\documentclass{article}

\newcommand{\uproman}[1]{\uppercase\expandafter{\romannumeral#1}}

\usepackage[english]{babel}
\usepackage{color}
\usepackage[a4paper,lmargin={2cm},rmargin={2cm},tmargin={2.5cm},bmargin = {2.5cm}]{geometry}
\usepackage{amssymb}
\usepackage{amsthm}
\usepackage{graphicx}
\usepackage{geometry}
\usepackage{amsmath}
\usepackage{xcolor}
\usepackage{listing}
\usepackage{wallpaper}
\usepackage{hyperref}
\usepackage{comment}
\usepackage[backend=biber, style= nature, sorting = none, maxbibnames=999]{biblatex}
\usepackage[font=footnotesize]{subcaption}
\usepackage{float}
\usepackage{stmaryrd}
\usepackage{setspace}
\usepackage{makecell}
\usepackage{acro}
\usepackage{array}
\usepackage{lineno}

\usepackage{enumitem}
\usepackage{tabto}

\usepackage{amsthm}
\usepackage{mathdots}

\usepackage{subcaption}
\newtheorem{theorem}{Theorem}

\theoremstyle{definition}
\newtheorem{definition}[theorem]{Definition}

\newcommand{\rgeq}{\mathbb{R}_{\geq 0}}
\newcommand{\R}{\mathbb{R}}

\DeclareAcronym{1}{
short = Mg,
long = Magnesium
}
\DeclareAcronym{2}{
short = ODE,
long = Ordinary differential equation
}
\DeclareAcronym{3}{
short = PDE,
long = Partial differential equation
}
\DeclareAcronym{4}{
short = Ti,
long = Titanium
}
\DeclareAcronym{5}{
short = Gd,
long = Gadolinium
}
\DeclareAcronym{6}{
short = BVTV,
long = Bone volume to total volume
}
\DeclareAcronym{7}{
short = Ca,
long = Calcium
}
\DeclareAcronym{8}{
short = BV,
long = Bone volume
}
\DeclareAcronym{9}{
short = TV,
long = Total volume
}
\DeclareAcronym{10}{
short = MAE,
long = Mean absolute error
}
\DeclareAcronym{11}{
short = NRMSE,
long = Normalized root mean square error
}
\DeclareAcronym{12}{
short = MOE,
long = Margin of error
}

\title{Computational modelling of bone growth and mineralization surrounding biodegradable Mg-based and permanent Ti implants}
\author{Nik Pohl$^{*,1}$, Domenik Priebe$^{*,1}$, Tamadur AlBaraghtheh$^{1,2}$, Sven Schimek$^{1}$, D.C. Florian Wieland$^{1}$,\\ Diana Krüger$^{1}$, Sascha Trostorff$^{3}$, Regine Willumeit-Römer$^{1,4}$,\\Ralf Köhl$^{3,4}$, Berit Zeller-Plumhoff$^{**,1,4,5,6}$}
\date{\small{*these authors, named alphabetically, have contributed equally to the manuscript}\\ \small{**corresponding author: berit.zeller-plumhoff@hereon.de, berit.zeller-plumhoff@uni-rostock.de}
\\ \small{$^1$ Institute of Metallic Biomaterials, Helmholtz-Zentrum Hereon, Germany}
\\ \small{$^2$ Institute of Surface Science, Helmholtz-Zentrum Hereon, Germany}
\\ \small{$^3$ Department of Mathematics, Faculty of Mathematics and Natural Sciences, Kiel University, Germany}
\\ \small{$^4$ Kiel, Nano, Surface, and Interface Science - KiNSIS, Kiel University, Germany}
\\ \small{$^5$ Data-driven Analysis and Design of Materials, Faculty of Mechanical Engineering and Marine Technologies, University of Rostock, Germany}
\\ \small{$^6$ Department Life, Light \& Matter University of Rostock, 18051 Rostock, Germany}
}

\begin{document}
\maketitle

\section*{Abstract}
\textit{In silico} testing of implant materials is a research area of high interest, as cost- and labour-intensive experiments may be omitted. However, assessing the tissue-material interaction mathematically and computationally can be very complex, in particular when functional, such as biodegradable, implant materials are investigated. In this work, we expand and refine suitable existing mathematical models of bone growth and magnesium-based implant degradation based on ordinary differential equations. We show that we can simulate the implant degradation, as well as the osseointegration in terms of relative bone volume fraction and changes in bone ultrastructure when applying the model to experimental data from titanium and magnesium-gadolinium implants for healing times up to 32 weeks. \textcolor{black}{An additional sensitivity analysis highlights important parameters and their interactions.} Moreover, we show that the model is predictive in terms of relative bone volume fraction with mean absolute errors below 6\%.

\section{Introduction}
The development of biodegradable implant materials to support bone healing temporarily is a highly active area of research with many advances in the experimental field. Magnesium (Mg)-based alloys have already been introduced to the market with first CE-certifications issued. Further developments are ongoing in optimizing degradation rates, mechanical properties and the tissue response \cite{TSAKIRIS20211884,HE20234396}. It is increasingly acknowledged that mathematical and computational models yield significant potential to accelerate the experimental work \textcolor{black}{in biomedical research, if predictive models can be achieved and implemented for \textit{in silico} experiments \cite{VICECONTI2021120,Myles_2023, Corral2020}}. However, such models will often result in high complexities, due to the large number of cell processes involved in bone healing and remodelling \cite{Borgiani2017, Checa2018, Kendall2023} and their interactions with the changing chemical environment during implant degradation. Therefore, simplified models that simulate the mechanistic behaviour of peri-implant bone remodelling and implant degradation \textcolor{black}{may be used in the absence of data to calibrate more complex models.} 

Previously, a number of models of bone healing and remodelling, and Mg alloy degradation have been introduced, see reviews by Carlier et al.~\cite{Carlier2015}, Wang et al.~\cite{Wang2017}, Oughmar et al.~\cite{Oughmar2020}, and AlBaraghtheh et al.~\cite{Albaraghtheh2022}. \textcolor{black}{Recently, \cite{Quinn2025} introduced a coupled model for bone fracture healing stabilized by Mg-based implants.} Many of these models are complex and defined using partial differential equations (PDEs) to incorporate spatial variations and dependencies. By contrast, Komarova et al.~\cite{Komarova2015} introduced a model of bone mineralization depending on inhibitors, nucleators for mineralization, and the formation of mature collagen matrices using ordinary differential equations (ODEs). The authors apply their model to pathologies in which bone mineralization is perturbed, including osteogenesis imperfecta, and are able to show the model's effectiveness in capturing main differences in the behaviour of the inhibitors and nucleators in these pathologies. 

In the case of modelling Mg degradation, a number of simplified models have been introduced. Grogan et al.~\cite{GROGAN-Acta} introduced an expression of the mass loss rate based on the diffusivity of Mg$^{2+}$ ions in the immersion medium and the saturation concentration of the ions therein. This simplified expression was able to yield results similar to a more complex PDE-based model. More recently, Gießgen et al.~\cite{Giessgen2019} presented a formulation of the degradation rate depending on the surface reaction rate and a parameter quantifying the amount of precipitated degradation product which hinders the degradation process. The authors highlight the improved performance of their model in capturing the experimentally observed degradation kinetics.

In the current work, we aim to use the model introduced by Komarova et al.~\cite{Komarova2015} to describe the bone remodelling process surrounding different implant types, specifically magnesium-gadolinium (Mg-xGd) and titanium (Ti) implants. We further extend the model by incorporating the effect of Mg alloy degradation on the bone remodelling process. Finally, we include details on the mineralization process, namely the individual hydroxyapatite crystal growth and the influence of the different implant types thereon. To model the degradation process we implement the existing models by Grogan et al.~\cite{GROGAN-Acta} and Gießgen et al.~\cite{Giessgen2019}. In doing so, we evaluate how modelling the degradation kinetics influences on the bone remodelling kinetics. The models are calibrated against data from experiments for which implant volume loss $V_{loss}$, peri-implant bone volume fraction $BV/TV$, and bone crystal lattice spacing $L$ and crystal width $C_{width}$ were quantified. The experiments to which we apply the models are those in which holes have been introduced into bone by drilling and an implant is then inserted using either a screw or a press fit. \textcolor{black}{In order to assess whether the coupled models are generalizable, we calibrate them to the data from Ti and Mg-10Gd implants. We then apply the fitted parameters to study how well the bone growth and remodelling around Mg-5Gd implants is described.}

\section{Mathematical formulation of models}
The ODE system used to describe Mg alloy degradation, bone growth and mineralization is derived in the following. A short explanation into the existence of solutions and their uniqueness is given in appendix \ref{sec:maths}.

\subsection{Modelling Mg alloy degradation}
The mathematical descriptions for the volume loss rate $\frac{dV_{loss}}{dt}$ are taken from the literature.
The diffusion-based power law introduced by Grogan et al.~\cite{GROGAN-Acta} for the mass loss rate is given as: 
\begin{align}
\frac{dM}{dt} &= \frac{\alpha D^\beta c_{sat}}{\sqrt{t}} \,,
\end{align}
where $D$ is the diffusion coefficient of Mg$^{2+}$ ions in the body fluid, $c_{sat}$ their saturation concentration and $\alpha, \beta > 0$ are constant coefficients. At constant temperature $D$ and $c_{sat}$ are constant and due to the proportionality of mass and volume, we reformulate the equation to describe $\frac{dV_{loss}}{dt}$ by introducing $m_1$:
\begin{align*}
    \frac{dV_{loss}}{dt} &= \frac{m_1}{\sqrt{t}} \tag{1$\alpha$}
\end{align*}

The alternative description of volume loss is based on Gießgen et al.~\cite{Giessgen2019}. The authors introduced the following equation to describe the degradation rate of a degradable metal:
\begin{align}
     \frac{dy_{corr}}{dt} &= \frac{r\cdot d}{rt+d}  \,,
\end{align}
where $y_{corr}$ is the average degradation depth given a surface degradation rate $r$ and a parameter $d$ that describes the influence of precipitated degradation products on the degradation process. The degradation rate is proportional to the volume loss rate by a factor $V_0/A$, where $V_0$ is the initial volume of the implant and $A$ its surface area. Thus, the volume loss rate can be described as: 
\begin{align*}
     \frac{dV_{loss}}{dt} &= \frac{A\cdot r\cdot d}{V_0(rt+d)} =  \frac{r'd'}{r't+d'} \tag{1$\beta$}
\end{align*}
for $r' = Ar/V_0, ~d' = Ad/V_0$.

\subsection{Modelling peri-implant bone formation}

The model of peri-implant bone growth and mineralization is based on the model presented by Komarova et al.~\cite{Komarova2015}. We utilize the model formulation that was non-dimensionalized spatially and formulated with respect to concentrations.
Komarova et al.~introduced their model of formation of mineralized bone based on the concept of bone tissue formation initiated by osteoblasts in the form of a naive collagen matrix $x_1$. The matrix transforms into mature collagen matrix $x_2$ at a characteristic rate $k_1$ over time. This matrix will be mineralized as the amorphous hydroxapatite located between the collagen fibres is taking crystalline form. This process is hindered by a combination of molecular inhibitors $I$ which diffuse through the naive collagen matrix at a rate $v_1$ and are removed during matrix maturation at a rate $r_1$ so that the mineralization can occur. \textcolor{black}{Komarova et al.~\cite{Komarova2015} thus incorporates the combined action of key inhibitors \cite{mckee2012bone}, i.e., pyrophosphate \cite{ORRISS201657}, and the small, integrin-binding ligand, N-linked glycoprotein (SIBLING) family of proteins \cite{staines2012importance} as a generalized inhibitor concentration I.} The mineralization of the hydroxyapatite begins at a number $k_2$ of nucleation sites $N$ in the gaps between collagen fibres. These sites are created during matrix maturation. Any nucleation site that is enclosed in mineral $H$ at a rate $r_2$ becomes inactive. The change in bone mineral depends on the inhibitors by means of a Hill function dependency, as well as the nucleation sites and a characteristic rate $k_3$.

In applying this model to describe the formation of bone surrounding permanent implants made of Ti, we assume that the processes occur similarly and any specific influence of the implant is negligible and can be accounted for by the existing parameters. This is a reasonable assumption as the implant is fitted into a drilled hole and it is not degradable. Thus, new bone may be assumed to form based on the process described by the model. Moreover, we equate the amount of formed mineral concentration $H$ to the amount of mineralized bone which is experimentally often quantified using the parameter of relative bone volume fraction $BV/TV$. We assume for our model that $H$ and $BV$ are proportional since the hydroxyapatite is assumed to grow homogeneously around the nucleators. Consequently, the amount of molecules may be assumed to be proportional to the volume in which they are growing, which again are homogeneously distributed in the bone volume and therefore the bone volume grows proportionally to the hydroxyapatite molecule number. Therefore, $H$ and $BV/TV$ are proportional to each other by further accounting for the constant total volume investigated. Thus, we may incorporate this proportionality constant in an updated parameter $k_3'$. 

In the case of Mg-based implants, the model additionally needs to account for the effect of the degradation of the implant. \textcolor{black}{For our model we assume that bone grows similarly around Ti and Mg with the only difference being the influence of released Mg$^{2+}$ ions on the system.} It has previously been shown Mg leads to a delay in hydroxyapatite formation \cite{Ding2014}. Thus, the rate of release of Mg$^{2+}$ ions, which can be assumed to be inverse to the rate of volume loss, influences the inhibitor concentration $I$ at a rate $m_2$. Therefore, the formation of bone around Ti and Mg alloy implants can be described by the following system of ODEs. The term added in gray is that added to the model introduced by Komarova et al.~\cite{Komarova2015} used only when describing bone formation around Mg alloys:

\begin{align*}
            \frac{dx_1}{dt} &= -k_1x_1\tag{1a}\\
            \frac{dx_2}{dt} &= k_1x_1\tag{1b}\\
            \frac{dI}{dt} &= v_1x_1 - r_1x_2I \textcolor{gray}{+m_2\frac{dV_{loss}}{dt} \tag{1c}}\\
            \frac{dN}{dt} &= k_2\frac{dx_2}{dt} -r_2\frac{dH}{dt}N\tag{1d}\\
            \frac{dH}{dt} &= k_3 \cdot \frac{b}{b+(I)^{a}}\cdot N \tag{1e}\\
\end{align*}
All parameters $k_1,v_1,r_1,m_2,k_2,r_2,k_3 \in \mathbb{R}_{>0}$. Because $V_{loss}$ is a fraction, no non-dimensionalization is necessary. 

\subsection{Modelling bone mineral growth}
To model the growth of the hydroxypatite crystals in more detail as this relates to the mechanical properties of the bone \cite{Iskhakova2024} and include the effect of the released Mg$^{2+}$ ions thereon, we further introduce models to describe the change in crystallite size and lattice spacing of the hydroxyapatite crystal over time. 

To this end, we assume that the hydroxyapatite nucleates as thin layers within the gaps between collagen fibers \cite{Fratzl1991}. Hydroxyapatite has a hexagonal shape and is assumed to grow first along its (002) plane before growing in width along the (310) plane \cite{Fratzl1991}. Thus, initially, the hydroxyapatite crystal can be described by a needle-like structure that later expands into platelets \cite{Xu2020}. Therefore, we assume a fixed crystal length, which has been shown to measure 30-50 nm \cite{Lotsari2018}. In accordance with the information provided by Ostapienko et al.~\cite{Ostapienko2019} for crystal growth, we model the change in crystal thickness $C_{width}$ as a function depending on the difference in hydroxyapatite concentration on the existing crystal surface and surrounding body fluid:

\begin{align}
    \frac{dC_{width}}{dt} = K_c L_{SA} (c_0 - c^*)^g \,,
\end{align}

where \(K_c\) is the crystallization reaction constant, \(L_{SA}\) is the surface area of the crystal, \(c_0\) is the solute concentration in the fluid, \(c^*\) is the equilibrium concentration and \(g\) represents the order of the crystal growth process.
As the crystal is assumed to grow over time, \(L_{SA}\) cannot be considered constant and should be approximated.

The surface area of an individual hydroxyapatite crystal needle/platelet is difficult to determine, therefore, we assume that the crystal geometry can be represented by a prism with a regular \(n\)-sided base, where \(n > 3\). Thus, we may approximate both the needle-like and platelet forms effectively. The following expression can be derived for the surface area, the derivation of which is given in appendix~\ref{sec:LSA_deriv}:

\begin{align}
    L_{SA} = k_{10} \cdot C_{width} \,,
\end{align}

with a proportionality constant $k_{10}\in \mathbb{R}_{>0}$.\\
The difference between the equilibrium concentration and the body fluid, \((c_0 - c^*)\), drives the crystal growth in this model. Hydroxyapatite precipitates if \((c_0 - c^*) > 0\) \cite{Ostapienko2019} and would therefore increase $H$. Thus, the change in bone mineral content \(\frac{dH}{dt}\) relates to these concentrations by some rate constant $k_5$: \((c_0 - c^*) = k_5 \frac{dH}{dt}\).  If \((c_0 - c^*) = 0\), then \(\frac{dH}{dt} = 0\). \textcolor{black}{This approach has the advantage that we do not need measurement data for $c_0$. Instead, we can use a previous ODE to adequately describe the process.} With \(k_6 := g\) and \textcolor{black}{\(k_4 := K_c \cdot k_{10}\cdot k_5^{k_6}\)}, the change in crystal width $C_{width}$ over time is described as:

\begin{align}
    \frac{dC_{width}}{dt} = k_4 \cdot C_{width} \cdot (\frac{dH}{dt})^{k_6}\,.
\end{align}

Finally, we aim to include the influence of Mg$^{2+}$ ion release on the crystal lattice spacing. This parameter is generally fixed for a bone with some spatial variations depending on health, age, etc. However, if certain ions are present in bone at sufficient concentrations, they may be incorporated into the crystal lattice by replacing Ca$^{2+}$ ions, thus changing the crystal lattice spacing due to a difference in atomic radius \cite{ZELLERPLUMHOFF2020,Bystrov2023}. 

Currently, no model has been published that considers a change in crystal lattice spacing due to a replacement of atoms over time. In this context, \(L_{Min}\) represents a theoretical minimal lattice spacing that could be achieved if all Ca$^{2+}$ were substituted with Mg$^{2+}$ in hydroxyapatite, while \(L_{Max}\) represents the lattice spacing for healthy hydroxyapatite crystals. However, replacing all Ca$^{2+}$ with Mg$^{2+}$ is not attainable in practice. Therefore, we reviewed existing literature to identify suitable values, 
which were taken as \(L_{Min}=3.403\,\text{\AA}\), \(L_{Max}=3.447\,\text{\AA}\) for the (002) plane and \(L_{Min}=2.243\,\text{\AA}\), \(L_{Max}=2.281\,\text{\AA}\) for the (310) plane \cite{Bystrov2023, GAYATHRI2018}.

Since the replacement of Ca$^{2+}$ with Mg$^{2+}$ at any time $t$ affects the lattice spacing, we consider that the reduction in lattice spacing occurs at a rate \(k_7\), which we assume to be proportional to \(\frac{dV_{loss}}{dt}\), until \(L_{Min}\) is reached. 
Therefore, our model of crystal lattice spacing change over time must include a term 
\begin{align}
    -k_7\cdot\frac{dV_{loss}}{dt}\cdot(L-L_{Min}) \,.
\end{align}

However, if the release of ions becomes negligible over time, the body can remove the incorporated ions and replace them with Ca$^{2+}$ ions instead. In that case, the lattice spacing increases again over time at a rate \(k_8\) until reaching the original lattice spacing \(L_{Max}\). In considering this, the crystal lattice change is affected by the following term: 
\begin{align}
    k_8\cdot(L_{Max}-L) \,.
\end{align}

Thus, we overall add two further equations to our ODE system in order to describe the crystal growth and change in lattice spacing over time:

\begin{align*}
            \frac{dC_{width}}{dt} &= k_4 \cdot C_{width} \cdot  (\frac{dH}{dt})^{k_6}\tag{1f}\\
            \frac{dL}{dt}&=-k_7\cdot\frac{dV_{loss}}{dt}\cdot(L-L_{Min})+k_8\cdot(L_{Max}-L)\tag{1g}\\
\end{align*}
with parameters $k_4,k_6,k_7,k_8 \in \mathbb{R}_{>0}$. $L$ is non-dimensionalized and normalized by $L^*=\frac{L}{L_{max}}$. Similarly, $K^*=\frac{K}{\hat{K}}$, where $\hat{K}= 15$\,nm \cite{In2020}, is the non-dimensionalized and normalized description of the crystallite size. 

\section{Materials and methods}
\subsection{Experimental data for model calibration}
The experimental data used for model calibration in this study stems mainly from animal experiments, in which Ti, Mg-5wt.\%Gd and Mg-10wt.\%Gd M2 screws have been implanted into the tibia diaphysis of Sprague Dawley rats. \textcolor{black}{The Mg-based materials were manufactured in-house, while Ti screws were obtained from Promimic AB (Mölndal, Sweden).} The animal experiments were conducted in two separate studies and they were performed after ethical approval by the ethical committee at the Malmö/Lund regional board for animal research, Swedish Board of Agriculture (approval number DNR M 188-15) and at the Molecular Imaging North Competence Centre Kiel (approval number V 241-26850/2017(74-6/17)), respectively. All experimental details have been published in \cite{KRUGER2022, IskhakovaCwieka2024, ZELLERPLUMHOFF2020}. The animals were sacrificed 4, 8, 10, 12, 20 and 32 weeks after implantation. At that point, the implant and bone surrounding it were explanted and studied using micro computed tomography. Based on the 3D images, which were evaluated at a voxel size of 5\,µm, the implant's volume loss (\%) and relative bone volume fraction (\%) in its vicinity were computed as described in \cite{KRUGER2022, IskhakovaCwieka2024}. Following cutting of thin sections of the explants, the bone ultrastructure was studied using X-ray diffraction. The crystal lattice spacing and crystallite size along the (310) direction, which is related to the crystal width, were computed for explants obtained after 4, 8 and 12 weeks \cite{ZELLERPLUMHOFF2020}. Moreover, the crystallite size and crystal lattice spacing along the (002) direction were computed for samples explanted 10, 20 and 32  weeks after implantation \cite{IskhakovaCwieka2024}. It was shown that the crystallite size along (002), which is related to the length of the crystal, did not differ significantly between time points. Thus, we model a change in crystal width only, but consider the (002) lattice spacing data in modelling.

\begin{figure}[htbp]
    \centering
    \includegraphics[width=\linewidth]{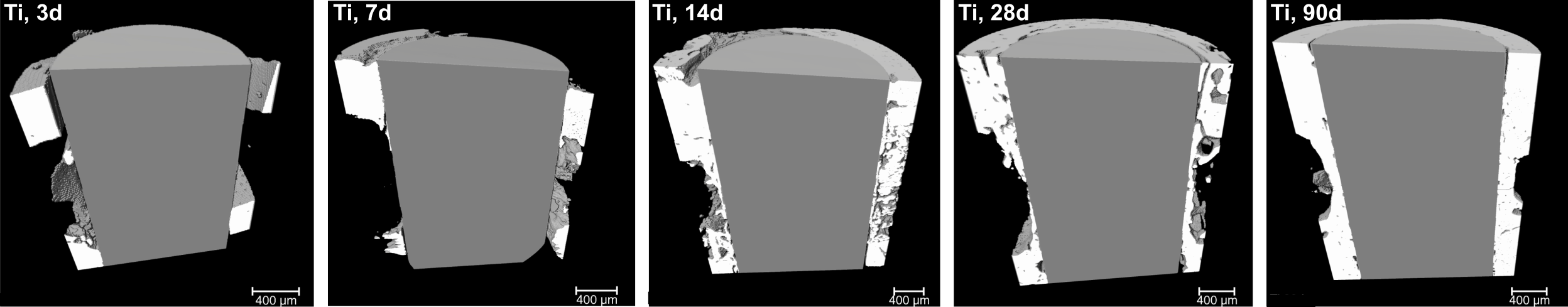}
    \caption{\textcolor{black}{3D renderings obtained after segmentation of the micro tomography image data for a representative sample from each time point. The renderings show the bone tissue (white) surrounding the Ti pin (grey) within the region of interest used for quantification of BVTV. All scale bars are 400\,µm.}}
    \label{fig:Ti_BVTV_Sven}
\end{figure}

In a second similar study, Ti pins with a diameter of 1.6\,mm and a length of\,8mm \textcolor{black}{produced from Ti6Al7Nb titanium (purchased from Medical University of Graz, Graz, Austria and manufactured by Acnis International, France)} were implanted into the femur of Sprague Dawley rats, which were sacrificed after 3, 7, 14, 28 and 90 days. The experiments were performed under ethical approval (Prot. n° 299/2020-PR) by the Instituto Superiore di Sanità on behalf of the Italian Ministry of Health and Ethical Panel. After making an insertion perpendicular to the femur shaft in the mid-diaphysis area. All skin layers were cut through and the femoris muscle tissues were carefully spread apart. After cleaning the bone of the residual soft tissue a hole with a diameter of 1.55\,mm was made. The sample was inserted by gentle tapping resulting in a press fit. Then the area was cleaned and the wound was closed. After the different healing times, the animals were sacrificed, and the implants with the surrounding bone were explanted. The bone-implant specimen were embedded in methyl methacrylate by the company LLS Rowiak LaserLabSolutions GmbH (Hanover, Germany). Micro computed tomography experiments were performed at the beamline P05 operated by Helmholtz-Zentrum Hereon at the PETRA III at Deutsches Elektronen Synchrotron (DESY, Hamburg, Germany). The measurements were carried out using a pixel size of 2.56\,µm and a resolution of\,2.85µm as determined by a mutual transfer function. Following tomographic reconstruction, the data was segmented using a nnU-net \cite{Isensee2021} as described in \cite{LopesMarinho2024}, and the relative bone volume fraction (\%) in the peri-implant region was computed. \textcolor{black}{Figure~\ref{fig:Ti_BVTV_Sven} shows volume renderings of the analysed volume of representative samples for each time point. Based on the image data, it is apparent that an increase in BVTV over time relates to the filling of gaps between implant and bone following implantation (days 3-14) and a subsequent formation of bone in the intramedullary region (days 14-90).}

\subsection{Initial values}
As initial conditions at $t=0$, we set $V=0$, as the implants are assumed to be non-degraded initially. Since equation 1$\alpha$ isn't defined for $t=0$, we are adding a small error $\varepsilon = 10^{-30}$ to $t$ to be able to solve the equation. We assume that $x_1=1$ since all matrix present is initially in its immature state and its production occurred sufficiently fast to be negligible to the model \cite{OrganicBoneMatrixByOsteoblasts}. Therefore $x_2=0$. In accordance with Komarova et al., we assume that $I=1$ and with no nucleators or mineralized bone present $N = 0$ \cite{Komarova2015}. Consequently, the hydroxyapatite crystal size $K=0$ at $t=0$ and we set the initial lattice spacing $L=L_{max}$, assuming an unperturbed crystal lattice. In the absence of data, we assume that $H(t=0)=0$. This enables us to still assess the differences between Ti and Mg-xGd implants, since it may be assumed that for similar implantation procedures, such as a press fit, the initial value for $H$ will be similar, and would only need to be offset by a constant. We perform a parameter study and vary the initial value when calibrating to the data from the first study to understand the impact of the initial value. Finally, to assess how well the model holds when $H(t=0)>>0$, we calibrate the model to the experimental data from the second study.

\subsection{Model implementation and calibration}

\textcolor{black}{The model equations are implemented in a Jupyter Notebook (Python 3.11.7) using the ODE-solver \textit{odeint} supplied in \textit{scipy} (v1.14.0) \cite{Virtanen2020}. The model calibration of $V_{loss}$, $H\propto BV/TV$, $C_{width}$ and $L$ is performed with respect to the median values from experiments. To this end, $L_2$ norm-based error functions are set up which are minimized using the covariance matrix adaptation evolution strategy (CMA-ES) available in Python within the package \textit{cma} \cite{hansen2019pycma}.} 

\textcolor{black}{We include an artifical data point at $t=3$ days with $H(t=3)=0$ for the calibration to the Ti-based implants in study 1. This additional point is ultimately required to achieve a sensible calibration of the model to the experimental data. The decision to set the artifical point at 3 days is based on fact that at such an early time point we can be more certain that the BVTV should remain unchanged from the initial state. This is based on two considerations: firstly, other processes occur in bone following implantation prior and in parallel to bone formation, such as the inflammatory response and angiogenesis. Initial bone formation during intramembraneous healing is visible around 7 days \cite{Vieira2015}. The second consideration for including an offset is based on the fact that bone mineralization starts between 5-10 days after matrix deposition \cite{MEUNIER1997373}. During the parameter study of $H(t=0)=0$, we are adjusting the value for the artifical time point accordingly. 
For the calibration of the model to data from Ti-based implants in study 1, it is further necessary to introduce regularization terms for $k_3$ and $v_1$. Thus, the loss function for Ti takes the form 
\begin{align}
\mathcal{L}=\frac{1}{n_{B}}\sum_{i=1}^{n_{B}} (BVTV(t_i) - H(t_i))^2 + \frac{1}{n_{C}}\sum_{i=1}^{n_{C}} (C_{width,exp}(t_i) - C_{width}(t_i))^2 + \lambda_k*{k_3}^2 + \lambda_v*{v_1}^2 \,,
\end{align} 
with regularization parameters $\lambda_k=0.001$ and $\lambda_v=0.01$ and $n$ representing the number of time points $t_i\,, i\in\{1,..,n\}$ at which experimental data is available for BVTV ($n_B=6$) and $C_{width}$ ($n_C=3$), respectively. The regularization parameters are set manually and they are required to avoid unrealistically large values for $k_3$ and $v_1$.}

Table~\ref{tab:par_ranges} contains the tested parameter ranges assessed for the different equations and respective variables and parameters. \textcolor{black}{To assess whether the CMA-ES algorithm may find more than one optimum depending on the initial parameter guess, the parameter initialization is performed using latin hypercube sampling (LHS) within the stated ranges. Based on 10 LHS samples, the optimal parameters for Ti and Mg-10Gd are determined. The algorithm is very stable, therefore, the search is restricted to one initialization for the calibration to Mg-5Gd. Since the calibration for $BVTV$ and $C_width$ for Ti and Mg-10Gd data is very demanding, the \textit{multiprocessing} library is used for parallelization.}

\begin{figure}[htpb!]
    \centering
    \includegraphics[width = 0.6\textwidth]{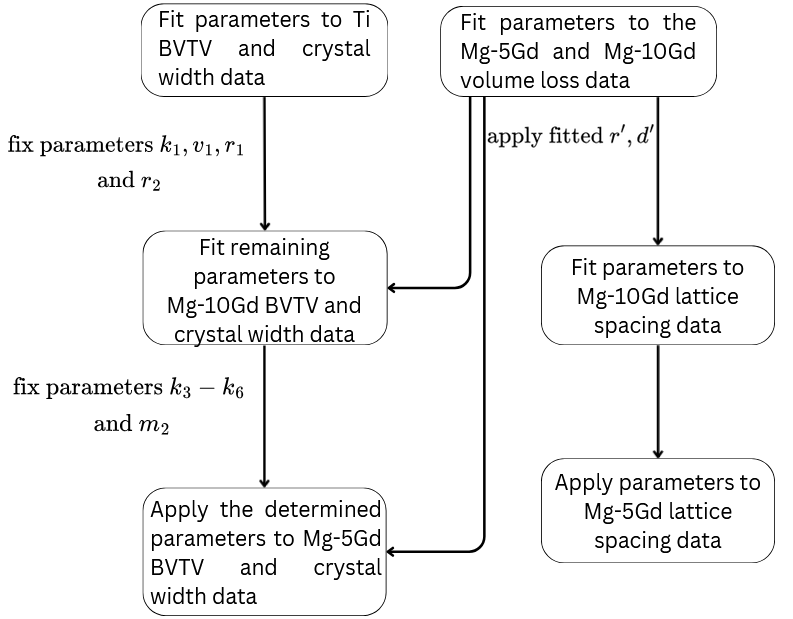}
    \caption{\textcolor{black}{Flowchart describing the order in which model parameters were fitted. Importantly, all parameters were fitted for Ti. At the same time VL is fitted for Mg-xGd implants. A number of parameters then fixed while remaining parameters are calibrated fit BVTV and $C_{width}$ for Mg-10Gd. Finally all parameters are fixed and the model is applied to the data of Mg-5Gd to assess the generalizability. Similarly, following fitting of parameters for volume loss, the lattice spacing parameters are fitted for Mg-10Gd and then applied to assess their generalizability for Mg-5Gd.}}
    \label{fig:flowchart}
\end{figure}

As stated above, we assume that the bone growth around Ti and Mg-based implants is generally similar, and that the main difference lies in the inhibition effect from the released ions. Therefore, we first calibrate all parameters for equations 1a-g to the data of the Ti screws, and set $m_2 = 0$ due to the lack of ion release. We then calibrate the only $m_2$ and $k_3$-$k_8$ to fit the equations to the data of Mg-10Gd screws. Finally, the same parameters are applied to the different degradation dynamics modelled for Mg-5Gd to assess the validity of the model. The parameters relating to the implant's volume loss are calibrated only in the case of Mg implants. We employ multi-objective optimization to simultaneously fit the parameters of equation $1f$ and $1e$. This approach has several benefits: it allows us to account for the interdependencies between the ODEs, enhances the overall accuracy of the parameter estimation and therefore provides a more comprehensive understanding of the system. \textcolor{black}{This workflow is visualized in Figure~\ref{fig:flowchart}.} 

$k_2=1$ was set for calibration to all materials based on the assumption that only one nucleator is present in the gap between collagen fibres. The Hill function coefficients were set to those values determined by Komarova et al., i.e. $a=10, b=0.001$. The order of crystal growth is $k_6 \geq 1$, with values of $k_6$ generally ranging between 1 and 2 \cite{MULLIN2001216} and experimental data indicating a parabolic dependence of hydroxyapatite precipitation on relative supersaturation \cite{Koutsoukos2022}. 

\begin{table}[htbp]
    \centering
    \begin{tabular}{|c|c|c|c|}
    \hline
        Equation &  Parameter  & Test parameter range & Physical meaning\\
        \hline
        1$\alpha$ & $m_1$ & [0, 1] & Scaled diffusion coefficient\\
        1$\beta$ & $r'$ & [0.0001, 1.2] & Surface corrosion rate*\\
        1$\beta$ & $d'$ & [0.0001, 1] & Fraction of corroded and precipitated material*\\
        1a,1b & $k_1$ & [0.01, 10] & Rate of collagen matrix maturation\\
        1c & $v_1$ & [0.001, 10] & Rate of inhibitor production\\
        1c & $r_1$ & [0.02, 20] & Rate of inhibitor removal \\
        1c & $m_2$ & [1, 150] & Effective rate of inhibition due to Mg degradation \\
        1d & $r_2$ & [1, 100] & Rate of encapsulation of nucleators\\
        1e & $k_3$ & Ti: [0.1, 10], Mg: [1, 30] & Bone growth rate\\
        1f & $k_4$ & Ti: [0.1, 100], Mg: [0.001, 300] & \makecell{Hydroxyapatite precipitation rate multiplied \\ with proportionality constant between crystal \\ width and surface area}\\
        1f & $k_6$ & Ti: [1, 3], Mg: [1, 5] & Order of precipitation process\\
        1g & $k_7$ & [0, 10] & Rate of Ca$^{2+}$ ion substitution\\
        1h & $k_8$ & [0, 1] & Rate of Mg$^{2+}$ ion removal from crystal\\
    \hline
    \end{tabular}
    \caption{\textcolor{black}{Model parameters and the parameter ranges tested for calibration. *corrected for the implant's initial surface area and volume.}}
    \label{tab:par_ranges}
\end{table}

Following optimization, the quality of the fit is determined using the mean absolute error (MAE) and the normalized root mean square error (NRMSE) computed based on all available data points. Moreover, for each model prediction we calculate the 95\,\% confidence interval by determining the model estimate at each time point and adding and subtracting the margin of error $MOE$. The margin of error was computed using the standard error of the model estimate and the critical value from the standard normal distribution corresponding to a 95\,\% confidence level \cite{hazra2017}:
\begin{align}
    MOE = 1.96 \times \frac{\sigma}{\sqrt{n}} \,,
\end{align}
$\sigma$ is the standard deviation of the model residuals, $n$ is the sample size (number of time points, at which experimental data is available) and 1.96 is the t-score for a 95\,\% confidence level. 

\textcolor{black}{\subsubsection{Sensitivity analysis}}
\textcolor{black}{To assess the influence of model parameters and their contributions to the variability of model outputs, we conduct a global sensitivity analysis using Sobol indices. Specifically, we calculate both first-order ($S_i$) and total-effect ($S_{T_i}$) Sobol indices for the key model parameters. $S_i$ (equation~\ref{eq:Si}) measures the individual contribution of parameter $i$ on the output variance, while $S_{T_i}$ (equation~\ref{eq:St}) captures the total contribution of the parameter, which includes both individual effects and interactions of the parameter with other parameters \cite{albaraghthehbest, Zhang2020ModernSampling}. 
\begin{equation}
S_i =  \frac{V_i}{Var (V)}  \quad ,
\label{eq:Si}
\end{equation} 
\begin{equation}
S_{T_i} =  S_i + \sum_{j\geq 1} S_{ij}+ ... + S_u = 1 - S_{\sim i} \quad ,
\label{eq:St}
\end{equation} 
where $V$, $V_i$ are the total variance of the model function and the partial variance contribution of each of the key parameters, respectively and $S_{\sim i}$ is the sum of all partial sensitivity indices $S_u$ except $S_i$. Sobol\textquotesingle~ indices are computed using SALib (Sensitivity Analysis Library in Python) \cite{herman2017salib, iwanaga2022toward}. SALib's saltelli sampling scheme is used to generate the necessary model evaluation points $2^{14}$ times. The tested parameters are assumed to follow uniform distributions within 20\% of the optimal parameter values determined during calibration. As the parameters relating to BVTV, $C_{width}$ and lattice spacing are fitted for Ti and Mg-10Gd and then applied to Mg-5Gd, the sensitivity analysis is performed for the former two materials only. For volume loss, the optimal parameter ranges of both Mg-5Gd and Mg-10Gd are considered for the sensitivity analysis.}
\section{Results and Discussion}

\subsection{Modelling Mg alloy degradation}

\begin{figure}[htpb!]
    \centering
    \begin{subfigure}{0.49\textwidth}
        \centering
        \includegraphics[width = \textwidth]{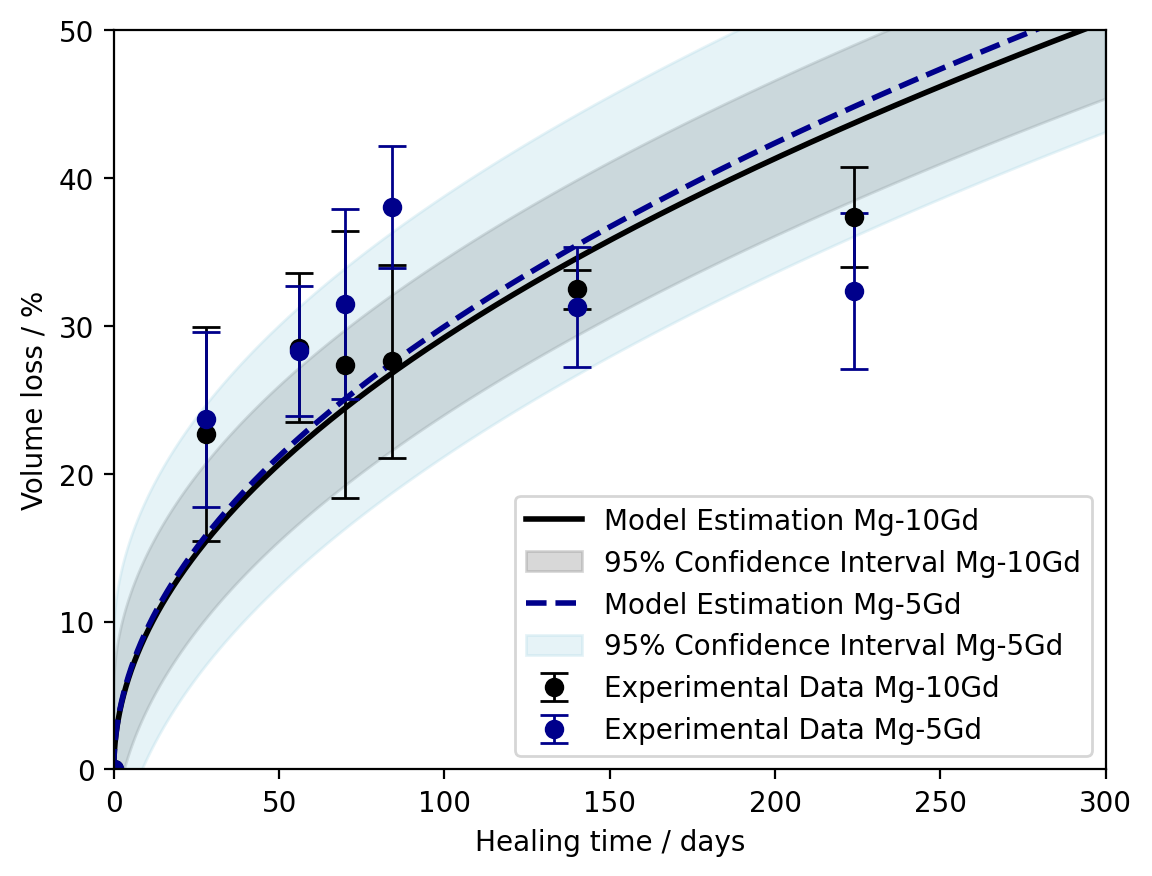}
        \caption{$V_{loss}$ modelled according to eqn.~$1\alpha$.}
        \label{fig:left}
    \end{subfigure}
    \begin{subfigure}{0.49\textwidth}
    \centering
        \includegraphics[width = \textwidth]{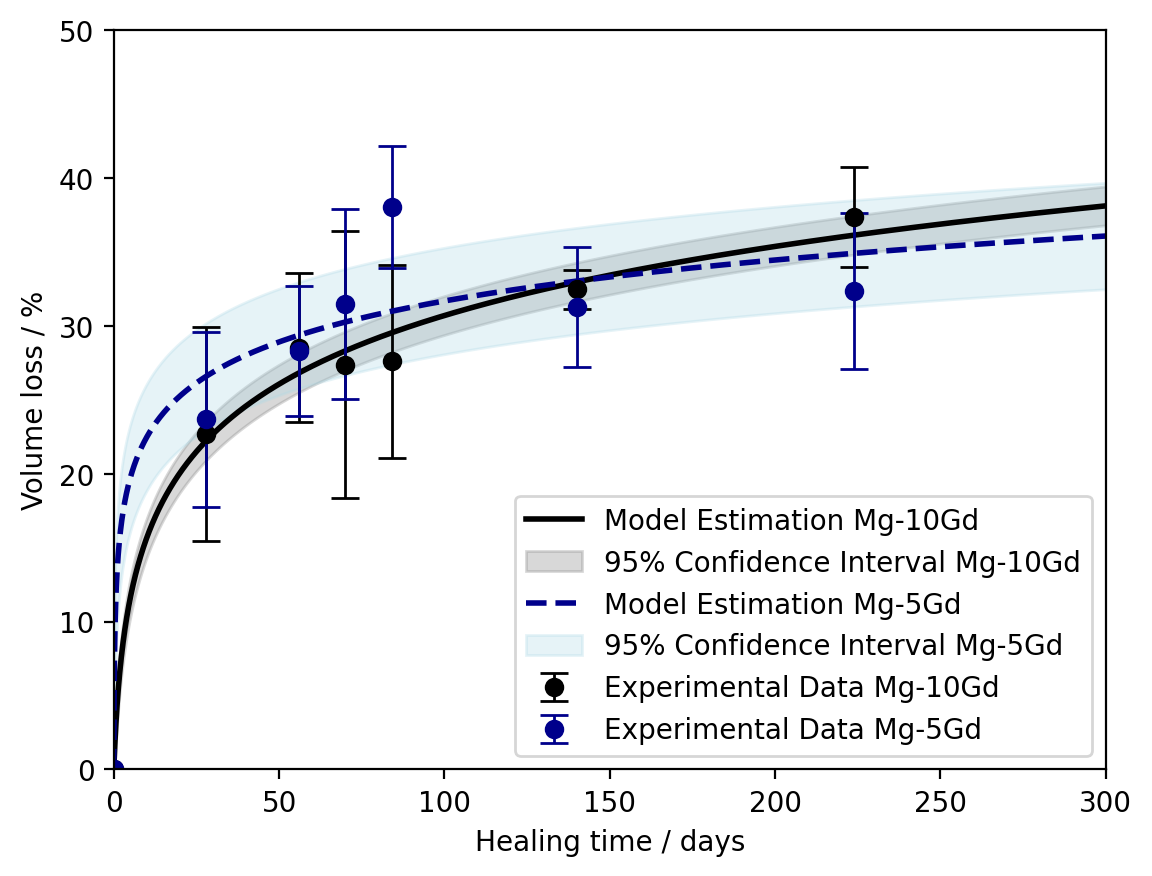}
        \caption{$V_{loss}$ modelled according to eqn.~$1\beta$.}
        \label{fig:right}
    \end{subfigure}
    \caption{Volume loss of Mg-5Gd (blue dashed line) and Mg-10Gd (black solid line) screws modelled according to eqns.~$1\alpha$ (a) and $1\beta$ (b), respectively. The experimental data was taken from \cite{KRUGER2022,IskhakovaCwieka2024} and is displayed as median and standard deviation. The model confidence intervals are indicated by transparent areas.}
    \label{fig:combinedVloss}
\end{figure}

Figure~\ref{fig:combinedVloss} shows the calibrated models for equations $1\alpha$ and $1\beta$ for the experimental data of Mg-5Gd and Mg-10Gd implants in study 1. It is visually apparent that the model presented by Gießgen et al.~\cite{Giessgen2019} captures the degradation dynamics more closely, while the power law model presented in \cite{GROGAN-Acta} overestimates the volume loss for later time points. \textcolor{black}{This is due to the fact, that the power law model does not take into effect the influence of the forming degradation layer on the corrosion dynamics. This effect is described explicitly by the model presented by Gießgen et al.~\cite{Giessgen2019} with the parameter $d'$.} There is no visual difference between Mg-5Gd and Mg-10Gd as modelled by equation $1\alpha$, while equation $1\beta$ indicates a different trend with higher initial volume losses for Mg-5Gd which level off more strongly over time. These trends agree with the published statistical analysis, which found no significant differences between the materials \cite{KRUGER2022,IskhakovaCwieka2024}. Moreover, \textit{in vitro} results had indicated that Mg-5Gd might degrade faster than Mg-10Gd at early time points \cite{KRUGER2021}, which would be supported by model $1\beta$. 

In accordance with the visual impression, the MAEs determined for model 1$\alpha$ are 7.90~\% for Mg-5Gd and 4.35~\% for Mg-10Gd, while that of model 1$\beta$ are 2.76~\% for Mg-5Gd and 1.12~\% for Mg-10Gd. The NRMSE values for all predictions are reported in table~\ref{tab:NRMSE} in and commented only when providing additional information to the MAE. The calibrated model parameters are given in table~\ref{tab:par_pot_VL}. These highlight the incapability of the power law to account for the difference in degradation dynamics displayed by Mg-5Gd and Mg-10Gd.

\begin{table}[htbp]
    \centering
    \begin{tabular}{|c|c|c|c|}
    \hline
        Equation &  Parameter  & Mg-5Gd & Mg-10Gd\\
        \hline
        1$\alpha$ & $m_1$ & 0.015 & 0.015\\
        1$\beta$ & $r'$ & 1.104 & 0.062\\
        1$\beta$ & $d'$ & 0.040 & 0.068\\
    \hline
    \end{tabular}
    \caption{Parameter values for models describing the volume loss calibrated to experimental data.}
    \label{tab:par_pot_VL}
\end{table}

 \subsection{Modelling bone growth}

\begin{figure}[H]
    \centering
        \begin{subfigure}{0.49\textwidth}
        \centering
        \includegraphics[width = \textwidth]{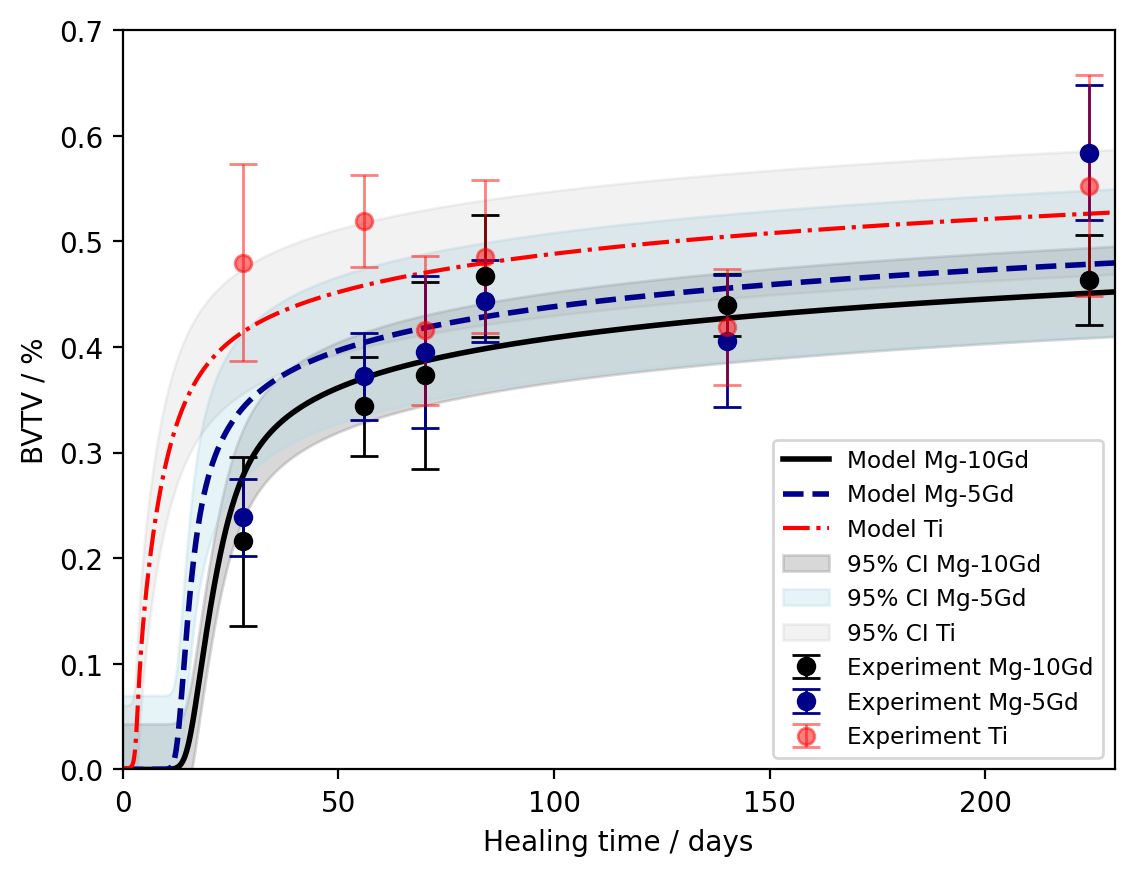}
        \caption{Simulation of BVTV corrected plot}
    \end{subfigure}
    \begin{subfigure}{0.49\textwidth}
    \centering
        \includegraphics[width = \textwidth]{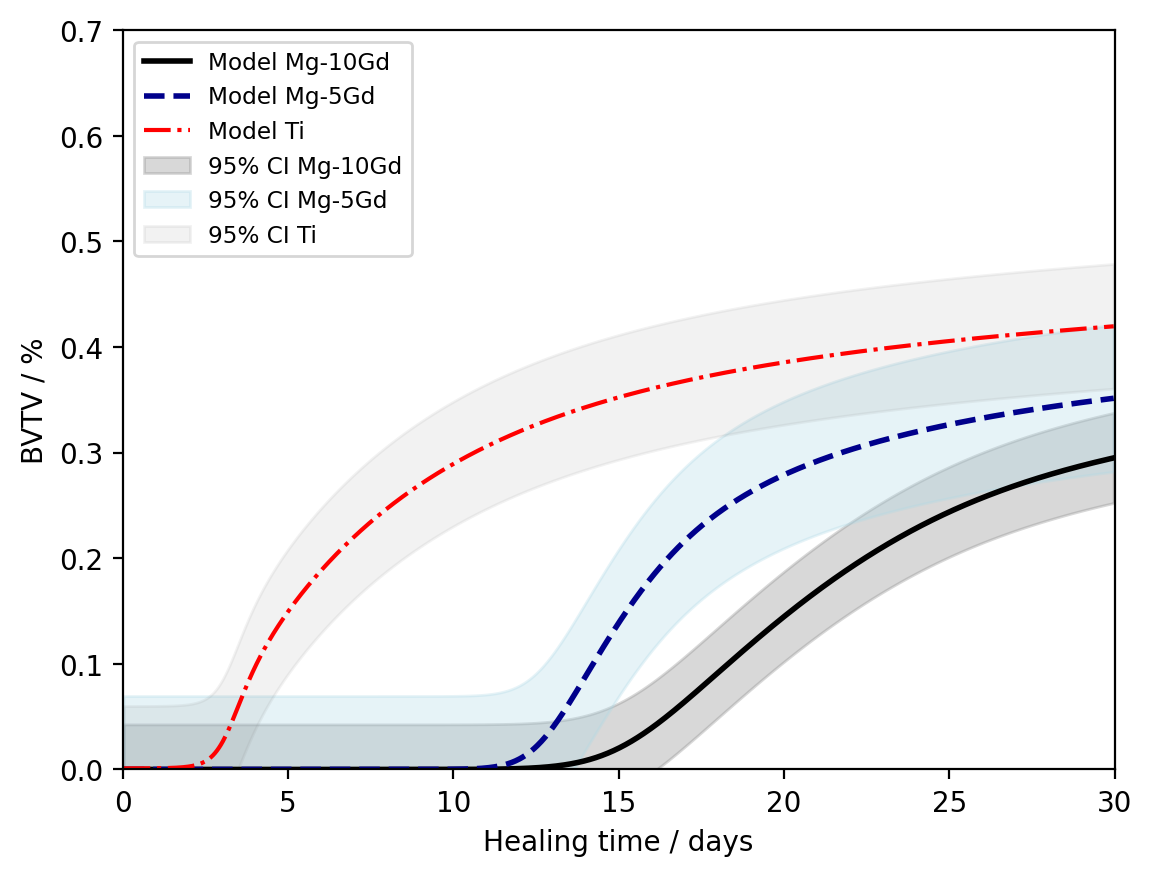}
        \caption{Zoom into (a) corrected}
    \end{subfigure}
    \caption{Simulated $BVTV$ according to equation 1e for Ti (red, dotted line), Mg-5Gd (blue, dashed line) and Mg-10Gd (black, solid line) implants. (a) shows the simulation results and experimental data over the course of 230 days, (b) is a zoom into the early time points. The experimental data was taken from \cite{KRUGER2022,IskhakovaCwieka2024} and is displayed as median and standard deviation. The model confidence intervals are indicated by transparent areas.}
    \label{fig:combinedBVTV}
\end{figure}

Because of the lower errors of model 1$\beta$ for simulating the volume loss, we will report the corresponding results for the modelled bone growth with respect to this model in the following. 
Figure~\ref{fig:combinedBVTV} shows the fitted $BVTV$ model for Ti, Mg-5Gd and Mg-10Gd implants.

The calibrated model parameters for both bone growth and mineralization are given in table~\ref{tab:par_pot_BVTV}. 

\begin{table}[htbp]
    \centering
    \begin{tabular}{|c|c|c|c|}
    \hline
        Equation &  Parameter  & Ti & Mg-10Gd\\
        \hline
        1a,1b & $k_1$ & 0.3880 & -\\
        1c & $v_1$ & 0.0164 & -\\
        1c & $r_1$ & 0.4915 & - \\
        1c & $m_2$ & - & 84.0455\\
        1d & $r_2$ & 23.1846 & -\\
        1e & $k_3$ & 0.5744 & 6.9291\\
        1f & $k_4$ & 38.1357 & 115.0688\\
        1f & $k_6$ & 1.2694 & 1.4712\\
        1g (310) & $k_7$ & - & 6.2619 \\
        1g (310) & $k_8$ & - & 0.0109 \\
        1g (002) & $k_7$ & - & 2.3037 \\
        1g (002) & $k_8$ & - & 0.0092 \\
    \hline
    \end{tabular}
    \caption{\textcolor{black}{Parameter values for equations 1a-1g after calibration to the experimental data. If no value is shown for Mg-10Gd, the value determined for Ti was set.}}
    \label{tab:par_pot_BVTV}
\end{table}

The MAE determined for the calibrated models shown in Figure~\ref{fig:combinedBVTV} is 4.98\% for Ti, 3.28\% for Mg-10Gd and 5.88\% for Mg-5Gd and confirms the visually apparent quality of the model fits. Importantly, this MAE is well below the variance present in the experimental data, highlighting the quality of the calibration. Moreover, the low error for Mg-5Gd shows the predictive power of the models, as no parameters where fitted in this case, expect for those describing $V_{loss}$. 

When considering Figure~\ref{fig:combinedBVTV}(b), it is apparent that the delay in the formation of bone depends on the implant material. In the case of Ti implants, the offset $t_{lag}$ was quantified to be around 9.6 days, which agrees with the literature \cite{Vieira2015}. The additional delay in bone formation for Mg-based implants relates to the influence of implant degradation. For the given models, based on equation 1$\beta$, this delay is higher for Mg-10Gd implants. This appears counter-intuitive, since a higher degradation rate would be associated with a higher release of ions which would hinder the mineralization, and bears further investigation.

\begin{figure}[H]
    \centering
    \begin{subfigure}{0.49\textwidth}
        \centering
        \includegraphics[width = \textwidth]{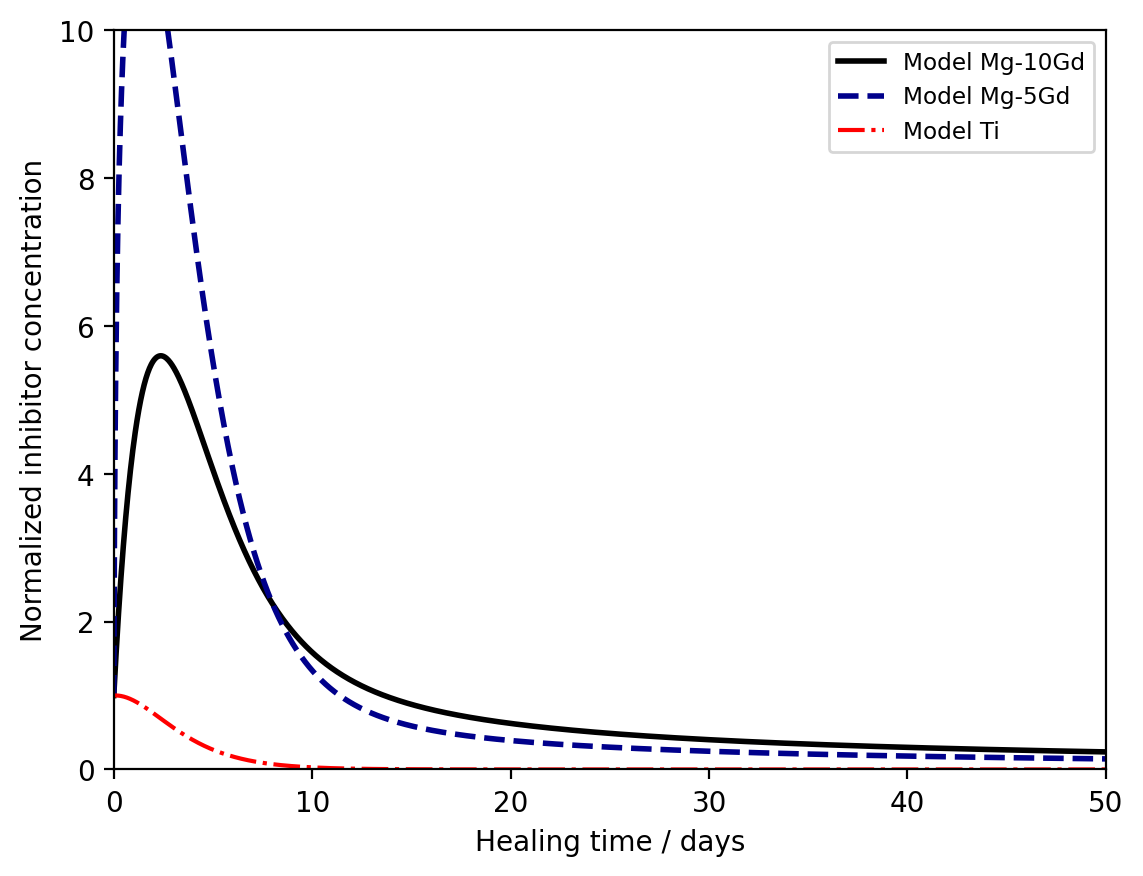}
        \caption{Normalized inhibitor concentration}
    \end{subfigure}
    \begin{subfigure}{0.49\textwidth}
    \centering
        \includegraphics[width = \textwidth]{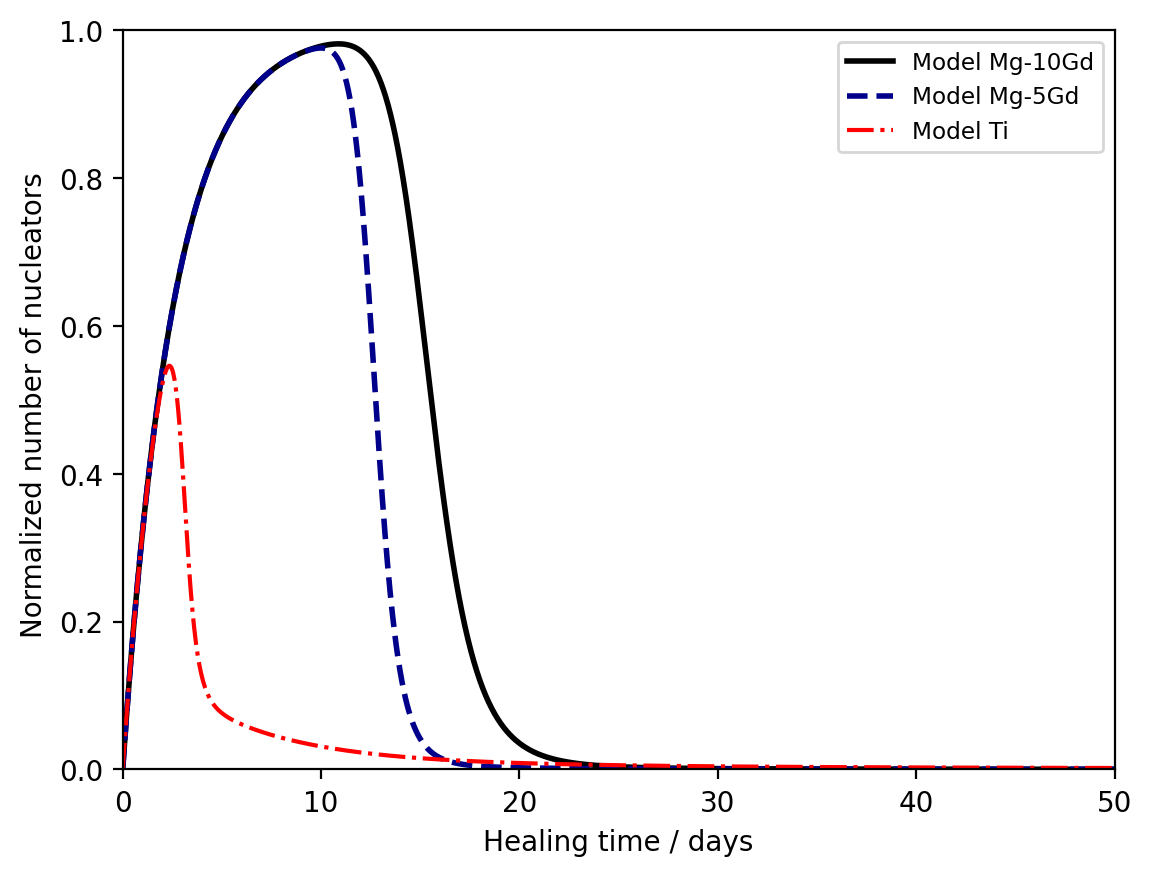}
        \caption{Normalized nucleator concentration}
    \end{subfigure}
    \caption{Model output for (a) the normalized inhibitor concentration $I$ computed according to equation 1c and (b) for the normalized nuclear concentration $N$ according to equation 1d for Ti (red, dotted line), Mg-5Gd (blue, dashed line) and Mg-10Gd (black, solid line) implants.}
    \label{fig:I_N_Giessgen}
\end{figure}

Considering the dynamics of inhibitor concentration in figure~\ref{fig:I_N_Giessgen}, it is apparent that $I$ was indeed highest for Mg-5Gd, until approx.~5 days. For both Mg implants, the inhibitor concentration was significantly higher than for Ti. Similarly, the nucleator concentration was higher for Mg-based implants, and decreased last for Mg-10Gd. The sustained high nucleator concentration can be explained by the high inhibitor concentration, resulting from our implementation where Mg acts as an inhibitor. This high inhibitor concentration prevents the mineralization of the nucleators. In our model, nucleators are only counted when they are not mineralized; once mineralization occurs, nucleator concentration decreases. Our Hill function is designed in such a way that mineralization only initiates when the inhibitor concentration is sufficiently low, leading to a high nucleator concentration as long as the inhibitor concentration remains high. Thus, because the inhibitor concentration degrades faster for Mg-5Gd than Mg-10Gd, bone formation is delayed for the latter.

\subsection{Modelling changes in hydroxyapatite}

\begin{figure}[H]
    \centering
    \begin{subfigure}{0.32\textwidth}
        \centering
        \includegraphics[width = \textwidth]{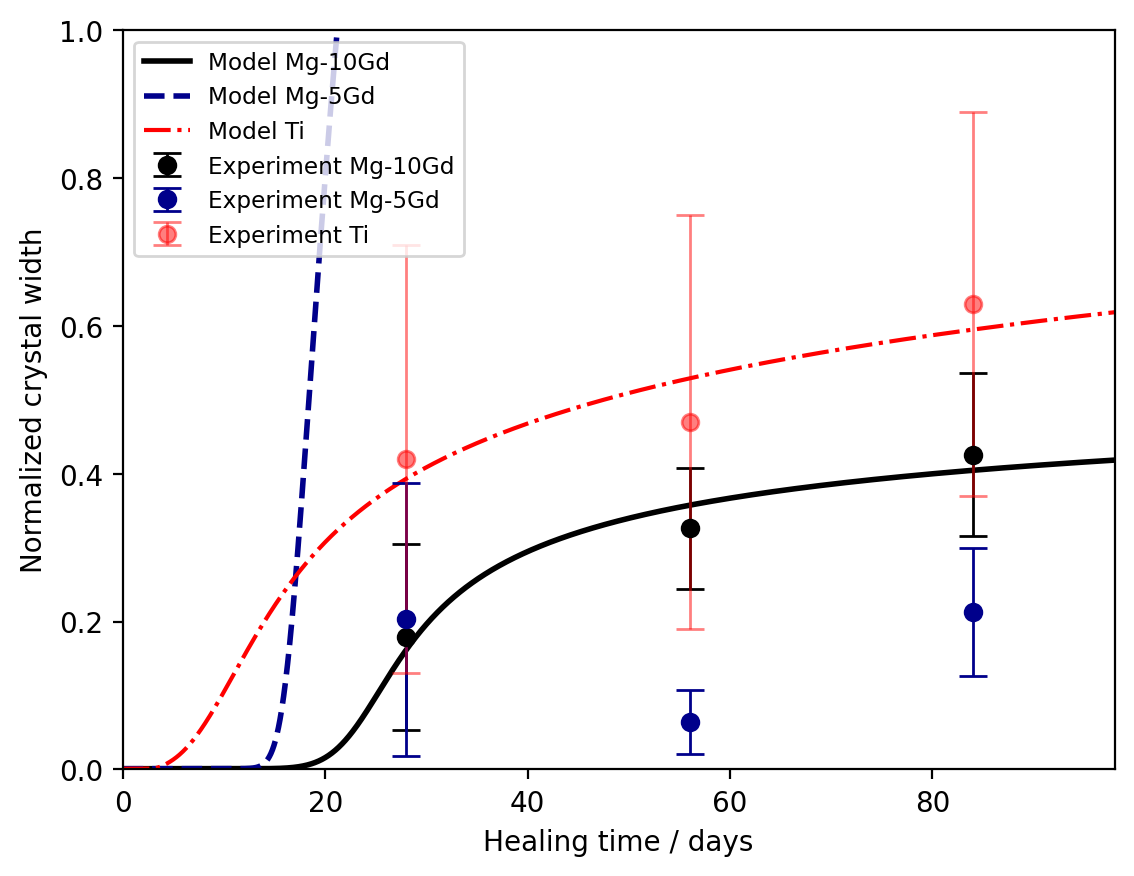}
        \caption{Crystal width $C_{width}$}
    \end{subfigure}
    \begin{subfigure}{0.32\textwidth}
    \centering
        \includegraphics[width = \textwidth]{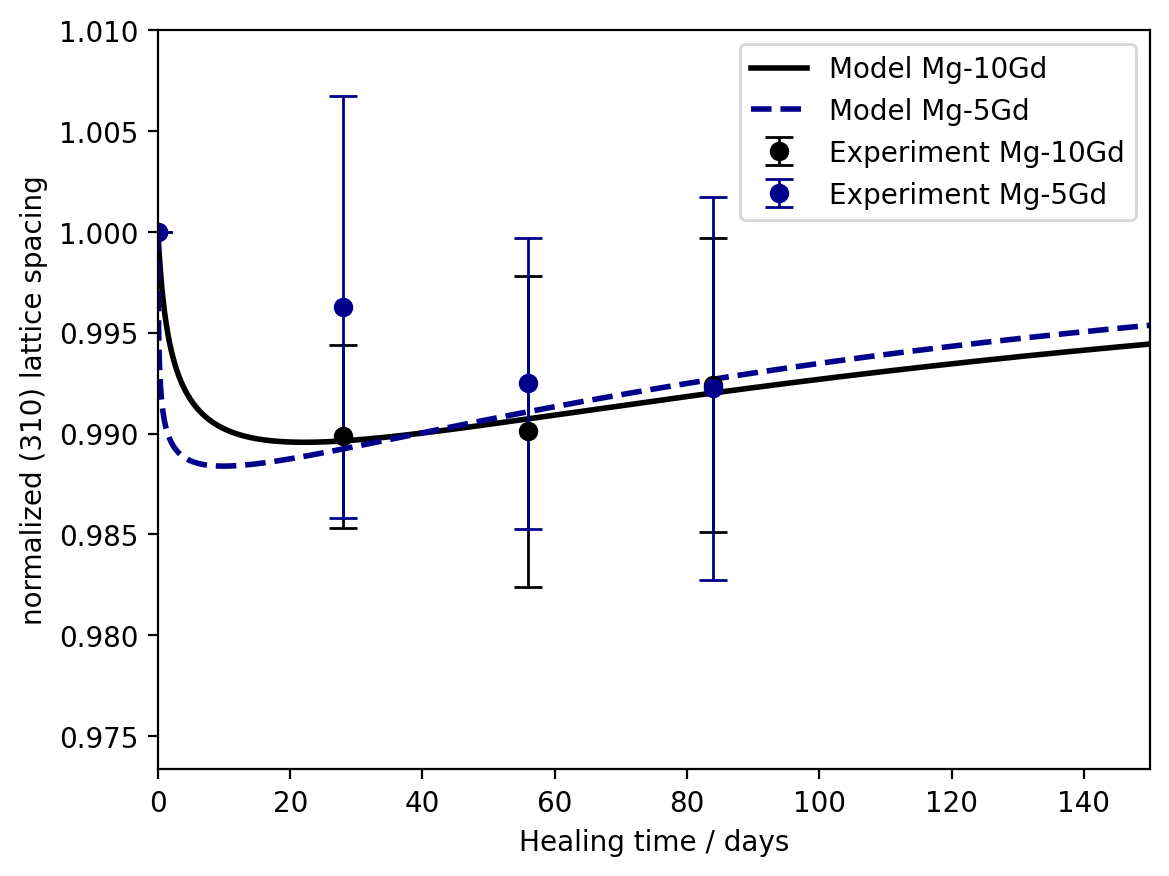}
        \caption{(310) lattice spacing $L_{310}$}
    \end{subfigure}
    \begin{subfigure}{0.32\textwidth}
    \centering
        \includegraphics[width = \textwidth]{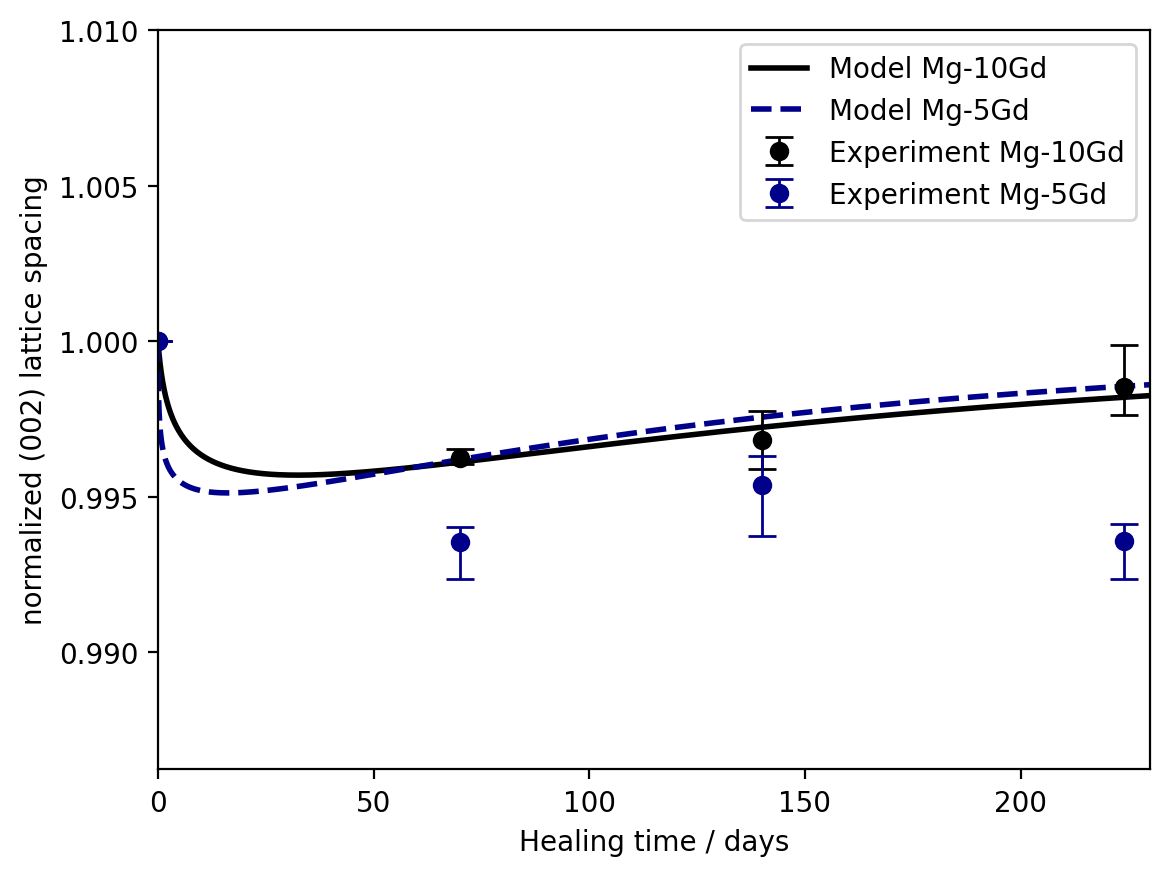}
        \caption{(002) lattice spacing $L_{002}$}
    \end{subfigure}
    \caption{Model output for (a) the normalized crystal width $C_{width}$ computed according to equation 1f  for Ti (red, dotted line) Mg-10Gd (black, solid line), and Mg-5Gd (blue, dashed line) implants and (b) for the normalized (310) and (c) the normalized (002) lattice spacing $L$ according to equation 1g for Mg-based implants. The experimental data in (b) is taken from \cite{ZELLERPLUMHOFF2020}, where it was given in mean $\pm$ standard deviation, while that in (c) is taken from \cite{IskhakovaCwieka2024}, which was given as median $\pm$ the 25/75th-percentile.}
    \label{fig:HAP}
\end{figure}

The changes in hydroxyapatite crystal width and lattice spacing according to equations 1f and 1g and calibration of the models to the data from study 1 are shown in Figure~\ref{fig:HAP}.

The resulting MAEs for the computed crystal width were 4.03~\% for Ti, 2.34~\% for Mg-10Gd and 2663.05~\% for Mg-5Gd, respectively. Similarly, and visually, the crystal width appears well calibrated for data of Ti and Mg-10Gd implants. Clearly, the fit for Mg-5Gd does not follow the data dynamics. This misfit is related to the high upper bound of $k_3$, which, if lowered to values below 10, will result in a fit of $C_{width}$ for Mg-5Gd closer to Mg-10Gd (not shown). Thus, we may need to consider deriving bounds based on physical meaning in the future. The fitted order of crystal growth $k_6$ is 1.09 and 1.25, respectively, which agrees with the literature \cite{MULLIN2001216}.

The prediction of the lattice spacings as displayed in Figure~\ref{fig:HAP}(b)-(c), shows that the equation 1g can be well calibrated to predict the change in lattice spacing for Mg-10Gd. When applying the fitted paramters for Mg-5Gd, however, the prediction appears less favourable. While the model fit is still within the reported standard deviation for $L_{310}$, this is not the case for $L_{002}$. The MAE for $L_{310}$ was computed as 0.0004 for Mg-10Gd and 0.0029 for Mg-5Gd, respectively. Similarly the MAE for $L_{002}$ was 0.0003 for Mg-10Gd and 0.0032 for Mg-5Gd, respectively. Due to the low absolute values of the lattice spacing the low MAEs may be misleading. The NRMSE reported in table~\ref{tab:NRMSE} highlights the fact that for $L_{002}$ in particular, the prediction of the model for Mg-5Gd is poor. This may partially be related to the data quality as standard deviations for the (310) reflection are very high. Similarly, the (002) data follows no observable trend and is significantly lower for Mg-5Gd than Mg-10Gd. It is possible, however, that parameters $k_7$ and $k_8$ are material specific since, e.g., $k_8$ effectively describes the removal of Mg$^{2+}$ ions from the bone crystal based on diffusive processes, and must therefore be calibrated for each material individually. In doing so, see appendix~\ref{sec:L_fitted}, the dynamics of crystal lattice change over time can be fitted for Mg-5Gd as well. Consequently, the model equation 1g appears well suited to describe the change in lattice spacing due the incorporation of Mg$^{2+}$ ions.

\begin{table}[htbp]
    \centering
    \begin{tabular}{|c|c|c|c|}
    \hline
        Variable &  Ti & Mg-5Gd & Mg-10Gd\\
        \hline
        $V_{loss}$ 1$\alpha$ & - & 0.58 & 0.34 \\
        $V_{loss}$ 1$\beta$ & - & 0.24 & 0.09 \\
        $BVTV$ & 0.41 & 0.38 & 0.28\\
        $C_{width}$ & 0.20 & 220.78 & 0.27\\
        $L_{310}$ & - & 1.04 & 0.18\\ 
        $L_{002}$ & - & 1.90 & 0.14\\
    \hline
    \end{tabular}
    \caption{\textcolor{black}{NRMSE values of predictions}}
    \label{tab:NRMSE}
\end{table}

\subsubsection{Dependence on initial value of $H$}
\begin{figure}[H]
    \centering
    \begin{subfigure}[t]{0.8\textwidth}
    \centering
        \includegraphics[width = \textwidth]{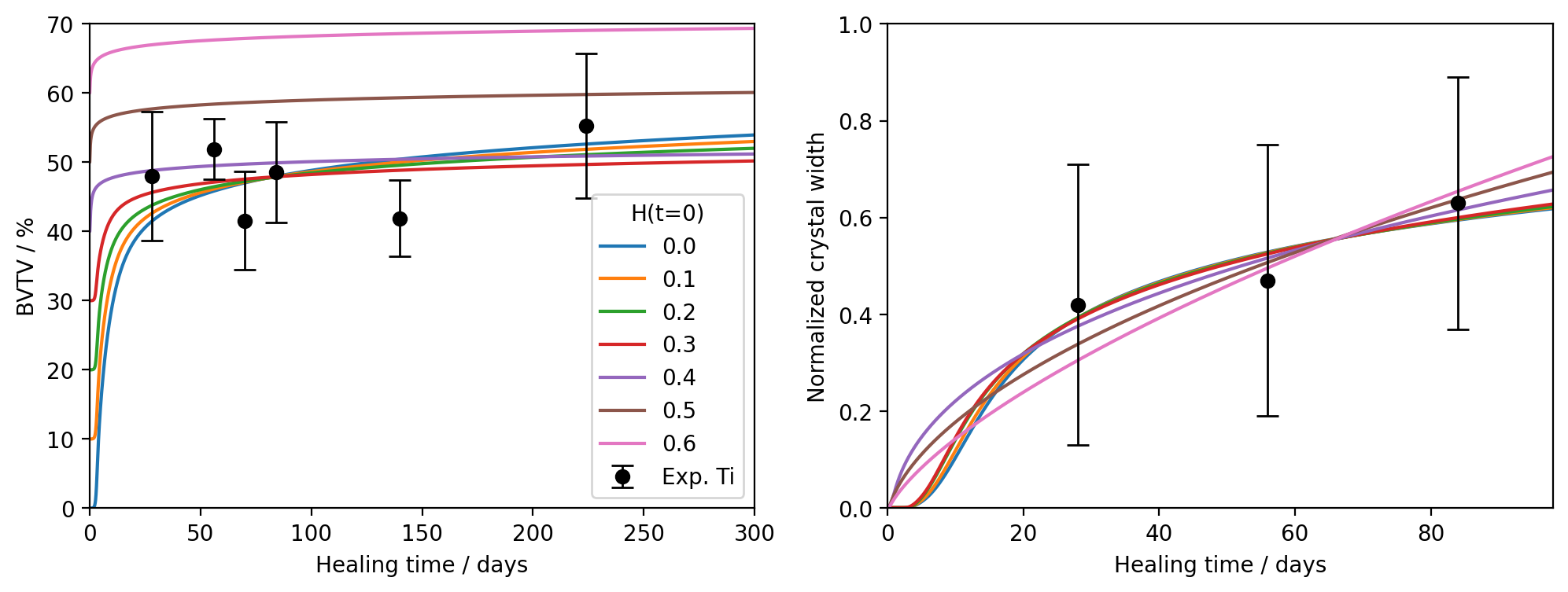}
        \caption{Fitted model curves for $BVTV$ and $C_{width}$ depending on different $H(t=0)$.}
    \end{subfigure}
    \begin{subfigure}[t]{0.4\textwidth}
        \centering
        \includegraphics[width = \textwidth]{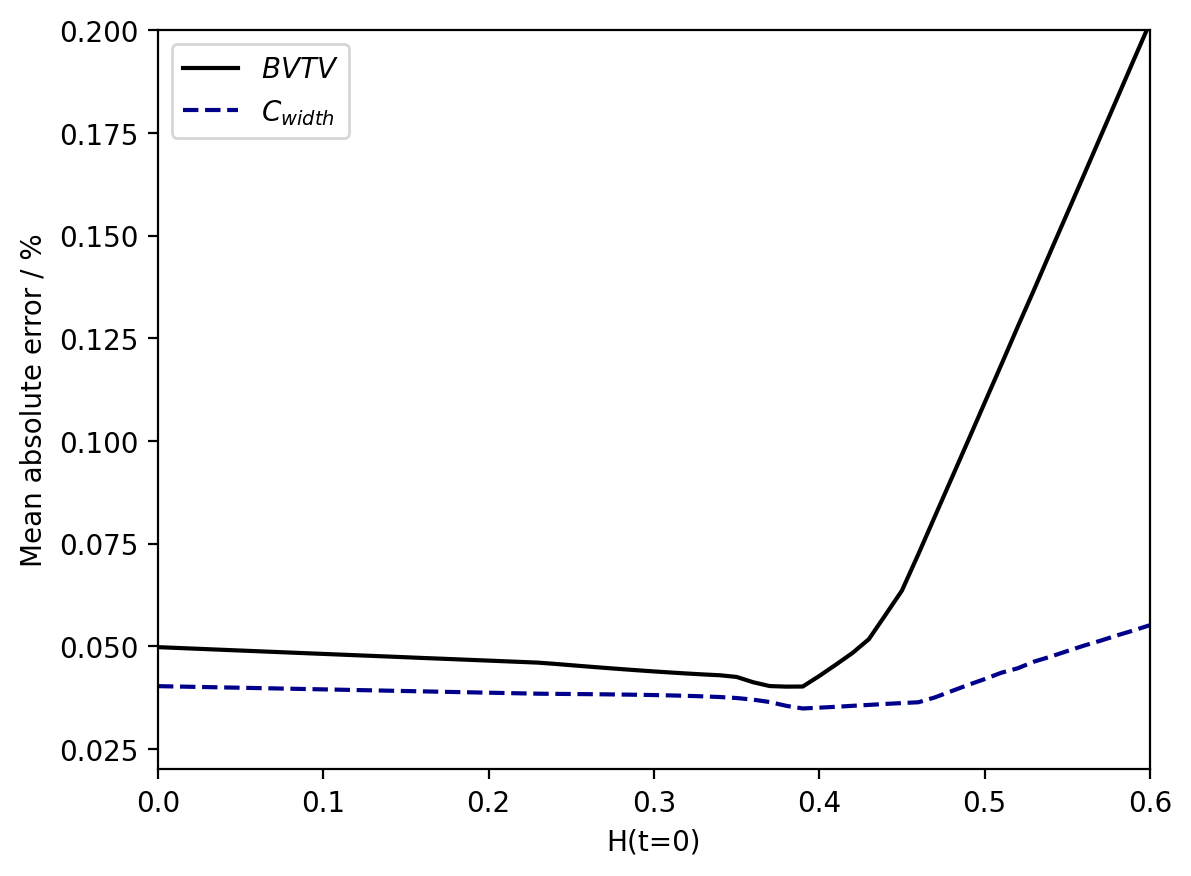}
        \caption{MAE as a function of $H(t=0)$ for calibration to Ti data from study 1.}
    \end{subfigure}
    \caption{Results of the parameter study for different $H(t=0)$ and calibration to the Ti data of study 1. (a) shows the resulting fitted curves for $BVTV$ and $C_{width}$ for a number of initial values and (b) displays the respective MAEs as a function of the initial value.}
    \label{fig:H_t0}
\end{figure}

To determine the influence of the initial value for $H$ in the absence of data for $BVTV$ at early time points, we have conducted a parameter study. The results are shown in Figure~\ref{fig:H_t0}. For \textcolor{black}{$H(t=0)\in [0,0.38]$, there is a slight decrease in the computed MAE for $BVTV$, due to the lack of experimental data at early time points and because higher values capture both the early and late time points well. Beyond that, the functions increasingly incapable to capture the experimental data} The MAE of the predicted $C_{width}$ similarly varies little. When considering the prediction, it is apparent that a change in $H(t=0)$ has little effect on the attainable quality of the fit. Therefore, we may conclude that the assumption of an $H(t=0)=0$ is acceptable in calibrating our model to the data from study 1. We must acknowledge, however, that any interpretation of model variables, such as $I$ and $N$ may be strongly influenced by this lack of data and knowledge. 

\begin{figure}[H]
    \centering
    \includegraphics[width = 0.5\textwidth]{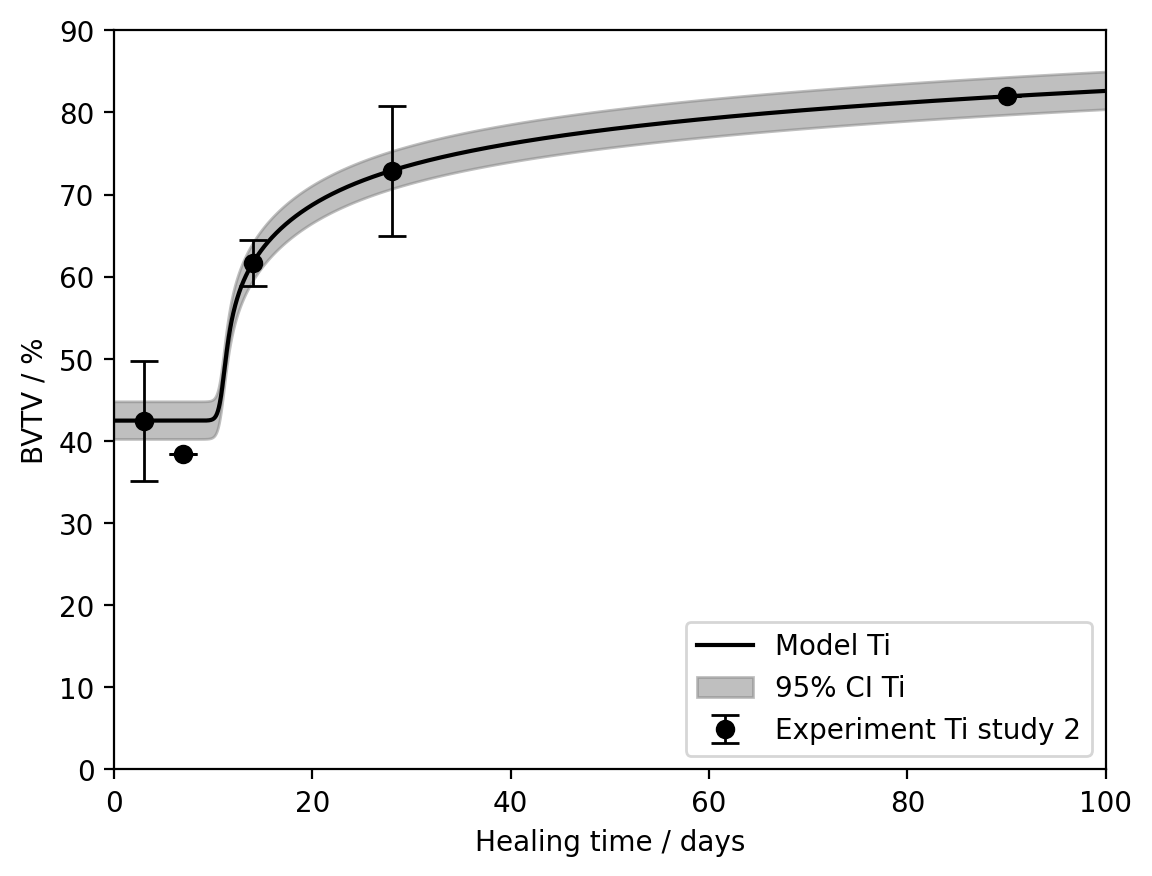}
    \caption{Simulated BVTV according to equation 1e for Ti pins from study 2. The experimental data is displayed as median and standard deviation.}
    \label{fig:BVTV_Sven}
\end{figure}

By contrast, earlier time points were investigated in study 2, including a 3-day time point. As indicated by the literature, no new bone growth would have appeared until this point, therefore, the median of the experimental data at this time \textcolor{black}{(42.47\,\%)} was set as the initial value for $H$. Figure~\ref{fig:BVTV_Sven} shows the resulting fitted curve after calibrating the same parameters as in study 1 for the case of Ti. The computed MAE was \textcolor{black}{0.82\,\%}. The quality of this fit highlights the suitability of the presented model to simulate peri-implant bone growth. \textcolor{black}{The corresponding fitted parameters are $v_1 = 0.3474,  r_1 = 9.6068, r_2 = 0.8024,  k1 = 18.5965,  k_3 = 0.6029$. A comparison with Table\ref{tab:par_pot_BVTV} indicates that except for $k_3$ all fitted parameters differ strongly between study 1 and study 2. To interpret this difference in more detail, more knowledge on the early time points in study 1 is required.} 

\begin{figure}[H]
    \centering
    \includegraphics[width = 0.8\textwidth]{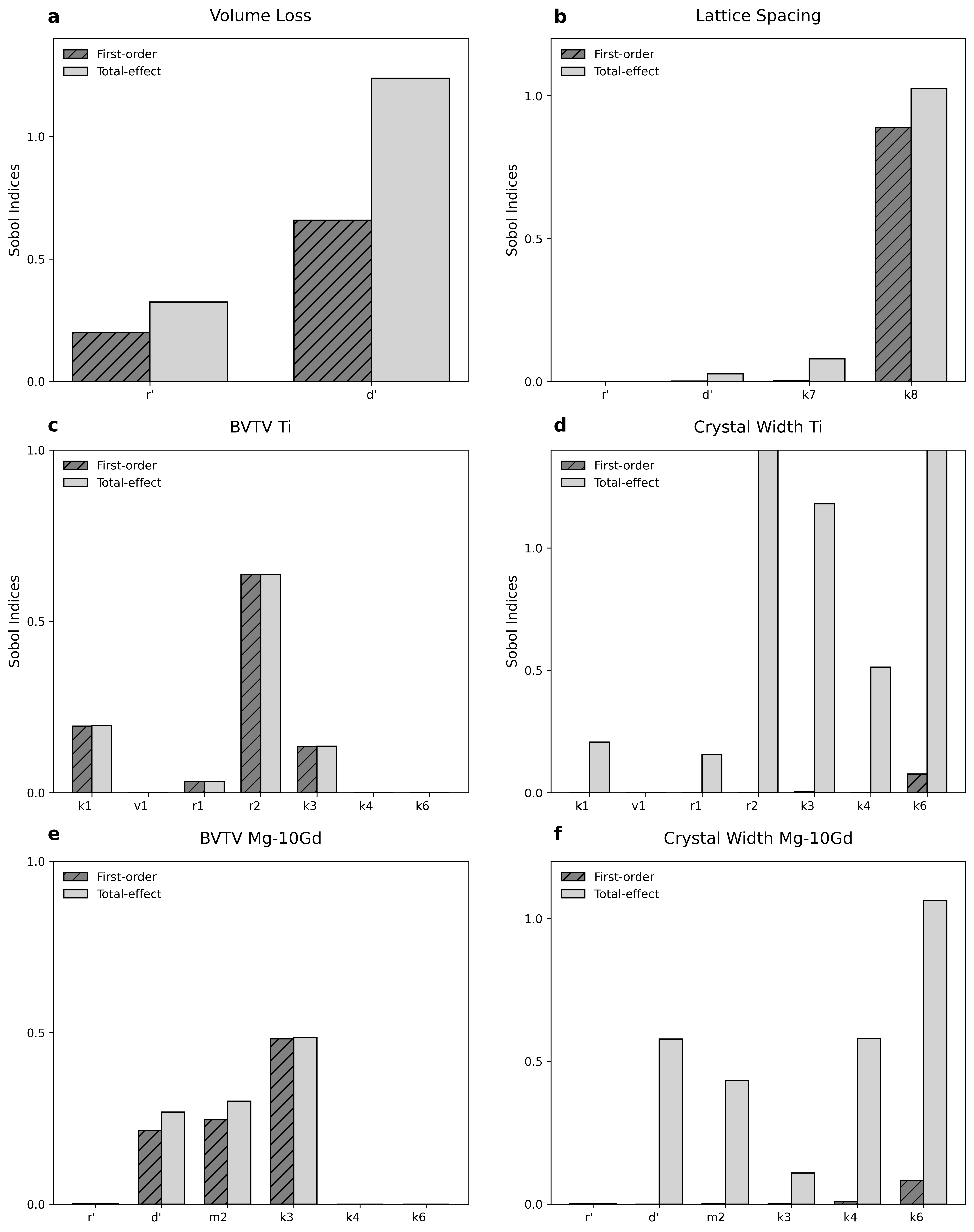}
    \caption{Global sensitivity analysis of mathematical models. The measured Sobol\textquotesingle~indices ($S_i$ and $S_{T_i}$) for (a) volume loss of Mg-xGd, (b) lattice spacing $L$ of Mg-10Gd, (c) BVTV of Ti (d) the crystal width of Ti, (e) BVTV of Mg-10Gd (f) the crystal width of Mg-10Gd.}
    \label{fig:SA}
\end{figure}

\textcolor{black}{\subsection{Sensitivity analysis}}
\textcolor{black}{Figure \ref{fig:SA} displays the results of the global sensitivity analysis, illustrating the first order ($S_i$) and total-effect ($S_{T_i}$) Sobol indices for each parameter in the model. Influential parameters are identified based on thresholds of $S_i \geq 0.05$ or $S_{T_i} \geq 0.1$. \\
We find that for volume loss (Fig.~\ref{fig:SA}a), parameter $d'$ has a notably higher individual effect than $r'$ with interaction between both. This indicates that the precipitation of degradation products on the implant surface has a major influence on the volume loss, which is in agreement with the literature. When assessing the sensitivity analysis for the lattice spacing (Fig.~\ref{fig:SA}b)) it is apparent that $k_8$ has the highest influence on the model output. $k_8$ corresponds to the rate at which Mg$^{2+}$ ions are removed from the hydroxyapatite lattice and replaced with Ca$^{2+}$ ions, thus indicating the importance of this mechanism.\\
For modelling BVTV with respect to data from Ti implants (Fig.~\ref{fig:SA}c), parameters $k_1$, $r_2$, and $k_3$ exhibit a high individual influence on variance, with $S_i$ values notably above the threshold level. $r_1$ shows a lesser influence.  Due to the similarity between $S_i$ and $S_{T_i}$ values, the interactions between variables appear neglible. Notably, the influence of $v_1$ is very low both for BVTV and crystal with. For the crystal growth model (Fig.~\ref{fig:SA}d), the individual effects of these same parameters are minimal, with $S_i$ values below 0.01, indicating negligible direct impacts on crystal growth predictions. $k_6$ has a somewhat larger effect, as it influences the power by which the change in hydroxyapatite formation influences the crystal width. The$S_{T_i}$ demonstrate strong correlations between all parameters (except $v_1$), particularly for parameters $r_2$, $k_3$, $k_4$, and $k_6$. This correlation implies that, while the parameters may not individually impact crystal growth, their interactions contribute significantly to the overall model variance, underscoring the importance of considering these interactions in the calibration process and performing a multiobjective optimization for calibration.
For the BVTV model applied to Mg-10Gd data (Fig.~\ref{fig:SA}e), the sensitivity analysis shows that parameter $d'$, $m_2$ and $k_3$ have the most significant effects. These parameters most strongly influence both the volume loss ($d'$), how it contributes as inhibitor ($m_2$) and how the inhibitors in turn affect the bone growth ($_k3)$. When considering their $S_{T_i}$ it is apparent that some interactions between $d'$ and $m_2$ will influence the overall bone growth. 
In terms of crystal width for Mg-10Gd (Fig.~\ref{fig:SA}f), parameter $k_6$ again stands out as the only parameter with notable indivial influence and with the highest $S_{T_i}$, indicating substantial interactions with other parameters and a major role in influencing the crystal growth variance for Mg-xGd. Similarly, for $d'$, $m_2$, $k_3$, and $k_4$ higher $S_{T_i}$ values emphasize the importance of their interactions within the model. This highlights the complex nature of the modelled crystal growth process and the influence of the implant's volume loss thereon, with the combined effects of multiple parameters shaping the overall model behaviour.}

\section{Conclusion and outlook}

By expanding and refining suitable existing mathematical models of bone growth and implant degradation, we are able to simulate implant degradation and osseointegration at micro- and ultrastructural level for Ti, Mg-10Gd and Mg-5Gd implants. Importantly, the predictive power of the models in modelling relative bone volume fraction is shown and we prove that the mid-term osseointegration response can be predicted well despite lacking data from early time points. Further testing of the model to simulate the response of the bone ultrastructure to biodegradable implants should be conducted in the future, to ensure the model's generalizability. \textcolor{black}{Additionally, the model should be expanded to include spatial dimensions so that the spatial variability and localized effects of implant degradation and tissue remodelling may be accounted for. Moreover, mechanical factors should be included to account for differences in load transfer between the implant types and corresponding mechanical stimulation of the bone remodelling. Thus, the short-, medium- and long-term stability of the implant, which is crucial for treatment success, may also be assessed.}

\subsection*{Acknowledgements}
The authors acknowledge funding from the German Ministry for Education and Research (OPTI-TRIALS - 16DKWN112C) and the NextGenerationEU program. We further would like to thank the DFG (Project number 526239533) for financial support. 
This research was supported in part through the Maxwell computational resources operated at Deutsches Elektronen-Synchrotron, Hamburg, Germany. 
The authors thank Hanna \'Cwieka and Kamila Iskhakova for their experimental work which provided data for model calibration in this work.


\subsection*{Data availability}
The data sets generated and analyzed during the current study are available from the corresponding authors on reasonable request.

\printbibliography

\appendix
\textcolor{black}{\printacronyms[heading= section, display=all,name={List of abbreviations}, template = tabular]}

\paragraph{}
\section{Mathematical analysis of existence and uniqueness of solutions}
\label{sec:maths}

When modelling real-life phenomena with ODEs there are several easy sanity checks one can do from a purely mathematical point of view in order to test the viability of the model: 
\begin{itemize}
    \item Does a (mathematical) solution of the ODE-model exist at all? If it does not, but apparently there is a solution in the real-life application, it is likely that the model is essentially flawed.
    \item Is the (mathematical) solution of the ODE-model unique? If it is not, it is possible that the simulation yield one of the solutions, while the real-life application would need a totally different solution to be treated and discussed. One should at least be careful, if non-uniqueness is an issue, and it is always easier to deal with ODE-models whose solutions are as unique as the real-life application requires.
    \item Is the solution positive? When measuring physical entities (like volume), a solution giving negative answers likely once again hints at a flawed ODE-model. (But also see the final bullet point for small volumes!)
    \item How stable is the solution under small perturbations of the initial data? Real-life measurement and numerical approximation are never precise. Therefore mathematical ODE-models should be forgiving to small perturbations of the initial data.
\end{itemize}

The following mathematical results help investigate these modelling questions.

\subsubsection*{Existence and uniqueness of solutions}

\begin{theorem}[{Local Picard-Lindelöf theorem \cite[pp.~416-417]{local-picard-Lindelöf}}]\label{local Picard-Lindelöf}
    Let $G\subseteq \R^2, (t_0,y_0)\in G$, such that for $\varepsilon, \delta > 0$ so that $U:=[t_0-\varepsilon, t_0+\varepsilon]\times [y_0-\delta, y_0+\delta]\subseteq G$ and $f: G\to \R$ continuous and locally Lipschitz-continuous with respect to the second variable. Let 
    \begin{align*}
        M := \sup_{(t,y)\in U}|f(t,y)|~, &&\alpha := \min\{\varepsilon, \frac{\delta}{M}\}
    \end{align*} 
    Then there exists exactly one solution to the initial value problem $\dot{y} = f(t,y)$ , $y(t_0) = y_0$ on the interval $[t_0-\alpha, t_0+\alpha]$.
\end{theorem}
Since all of our differential equations are characterized by using continuous and locally Lipschitz-continuous functions, this theorem tells us that there exist in fact unambiguous functions solving these differential equations for any initial values and times, at least for short time-frames. If we modify $(1d)$ by inserting the needed derivatives and solutions
\begin{align}
    \frac{dN}{dt} = k_2\frac{dx_2}{dt} -r_2\frac{dH}{dt}N\to \frac{dN}{dt} = k_1k_2x_1 -r_2k_3(\frac{b}{b+I^a})N^2
\end{align}
this equation becomes solvable. It is theoretically possible to then step by step construct a solution for any time frame by glueing local solutions together (this can also be done for the other ODEs, but it's simply easier and faster to use the next theorem). 

\begin{theorem}[{Global Picard-Lindelöf theorem \cite[p.~149, Theorem III.2.5.]{global-picard-Linedlöf}}]   \label{global Picard-Lindelöf}
    Let $G = [a,b]\times \R^n$, for $a,b\in \R$, $f:G\to\R^n$ a continuous function which satisfies global Lipschitz-continuity with respect to its second variable.
    
    Then for all $(t_0,y_0)\in G$ the initial value problem $\dot{y} = f(t,y)$, $y(t_0) = y_0$ is uniquely solvable on $[a,b]$.
\end{theorem}
It is fairly easy to show that the functions describing the other ODEs meet the requirements of the Global Picard-Lindelöf theorem, thus we can conclude that there exist explicit unambiguous functions solving these differential equations.

In most cases however, with the exception of $(1\alpha), (1\beta), (1a), (1b)$, these solutions cannot be expressed using elementary functions only. This means we still need to numerically solve ODEs or integrate to evaluate them.

\subsubsection*{Positivity of solutions}

\begin{theorem}[Additive Estimation]
    Let $T, Y\subseteq \mathbb{R}$ be intervals, let $f_+: T\to \mathbb{R}_{>0}$ and $g: T\times Y \to \mathbb{R}$ be continuous. If there are differentiable solutions $y$ and $y_h$ for the ODEs
    \begin{align*}
        \dot{y} = f_+(t) + g(t,y), && \dot{y_h}= g(t,y_h)
    \end{align*}on T, the following applies:\\
    If there exists $s\in T: y(s) > y_h(s)$, then $$\forall t \in T_{\geq s}: y(t)>y_h(t).$$
\end{theorem}
Since we cannot easily express solutions to our ODEs, we developed the additive estimation theorem which provides us with lower bound estimations through solutions of the homogeneous equations. Since these are comparably easy to solve and actually provide us with an expressable explicit function for $(1c)$ which stays positive for positive starting values, it automatically becomes clear that our desired solution behaves accordingly. The same can be said about solutions to $(1\alpha), (1\beta), (1a), (1b)$ just by looking at the explicit solutions for the initial value problems (IVPs) $t_0 = 0$ and $V_0, C_1, x_0 \geq 0 $:
\begin{align*}
    V_{loss}(t) &= V_0 + 2m_1 \sqrt{t}                   \tag{$1\alpha$}\\
    V_{loss}(t) &= V_0 + d\log(\frac{r't+d'}{d'})     \tag{$1\beta$}\\
    x_1(t) &= C_1e^{-k_1t}                                   \tag{1a}\\
    x_2(t) &=  x_0 + C_1(1-e^{-k_1 t})   \tag{1b}
\end{align*}

\begin{theorem}[Multiplicative Estimation]
    Let $T, Y\subseteq \mathbb{R}$ be intervals, let $f, h: T\to \R$ be continuous with $\forall t\in T: f(t)>h(t)$ and let $g_+: T\times Y \to \mathbb{R}_{>0}$ be continuous. If there are differentiable Solutions $y_f, y_h$ for the ODEs 
    \begin{align*}
        \dot{y_f} = f(t) \cdot g_+(t,y_f), && \dot{y_h}= h(t) \cdot g_+(t,y_h)
    \end{align*}
    on T, the following applies: if there exists $s\in T: y_f(s) > y_h(s)$, then $\forall t \in T_{\geq s}: y_f(t)>y_h(t)$.
\end{theorem}

This theorem was also developed to be used in tandem with additive estimation and essentially lets us do a lower bound estimation for $(1d)$ with a much simpler equation 
\begin{align}
    \frac{dN'}{dt} = r_2k_3{N'}^2.
\end{align} 
This equation is solvable and provides a positive expressable solution for positive initial values and thus the same can be said about the solution to the original ODE. Since $(1e)$ is mostly characterized by $N$, a positive solution $N$ already implies, that any solutions $y$ to a positive IVP for (1e) must already be positive. The same result follows analogously for $(1f)$ and $(1g)$, with $H$ and $V_{loss}$. 

\subsubsection*{Continuous dependency of the solution on initial data}

\begin{definition}[{Continuous Dependency \cite[p.~67, Definition 4.1.1.]{continuous-dependency}}]
    Let $G\subseteq \R\times\R^n$ be open, $f:G\to\R^n, (t,x)\mapsto f(t,x)$ continuous and locally Lipschitz-continuous with respect to the second variable. Let then $x$ be a solution to the IVP $(t_0,x_0)\in G$ and $\dot{x} = f(t,x)$ and $x(t_0) = x_0$ on its maximal interval of existence $T$. Let $I\subseteq T$ be compact and define $$graph_I(x) :=\{(t,x(t)): t\in I\}.$$
    $x$ is called \textbf{continuously dependent} of $(t_0,x_0,f)$ if for every compact interval $I\subseteq T$ there exists a compact neighborhood $K\subseteq G$ of $graph_J(x)$, so that the following holds:
    
    For every $\varepsilon>0$, there exists a $\delta >0$, so that for any $g\in C(K,\R^n)$  and $(\tau_0,y_0)\in K$ which satisfy 
    $$|t_0-\tau_0|<\delta, |x_0-y_0|<\delta, \sup_{(s,z)\in K)} |f(s,z)-g(s,z)|<\delta$$
    any solution $y$ which solves the IVP $(\tau_0,y_0)$ for $\dot{y}= g(t,y)$ exists on $I$ and $$\forall t\in I: |x(t)-y(t)|<\varepsilon.$$
\end{definition}

\begin{theorem}[{\cite[p.~68, Theorem 4.1.2.]{continuous-dependency}}]
    Let $G\subseteq\R\times\R^n$ be open with $(t_0,x_0)\in G, f:G\to\R^n$ continuous and locally Lipschitz-continuous with respect to its second variable. If $x$ is a unique non-extendable solution to the IVP $\dot{x} = f(t,x)$, $x$ is continuously dependent on $(t_0,x_0,f)$.
\end{theorem}
As discussed earlier we can find unique global solutions to every equation for positive IVPs, thus we can construct a globally unambiguous solution $X$ for the whole system. Continuous dependency essentially tells us that we can introduce small errors into the system, through for example different starting values or variables used, and still get similar behavior without extreme fluctuation. Since each equation is characterized by a locally Lipschitz-continuous function, the same holds true for the function $f$ characterizing the whole system. The Lipschitz-constant can be used to estimate this behaviour: For $t\in [t_{min},t_{max}]\subset \rgeq$, $\delta>0$, $Y:= (V,x_1,x_2,I,N,H,C,L)\in \rgeq^8$ and any $Y:= (V',x_1',x_2',I',N',H',C',L')\in \rgeq^8$ so that $||Y-Y'||_2\leq \delta$ it can be characterized by determining the following constants:

\begin{align}
K_1 :=& 2k_1+v_1+k_2k_3\\
K_2 :=& r_1(I+\delta)\\
K_3 :=&r_1x_1 + r_2k_3(N+\delta)^2 \hat{G}+ k_3(N+\delta)\hat{G}+ k_3^{k_6}k_4k_5^{k_6}k_6(C+\delta)\hat{G}\\
K_4 :=&r_2k_3(2N+\delta)+ k_3+  k_3^{k_6}k_4k_5^{k_6}k_6(C+\delta)\\
K_5 :=&\hat{\dot{V}}_{loss}\cdot k_7+k_8
\end{align}
Where 

\begin{align}
\hat{G}:= ab(I+\delta)^{a-1} + \ln (b+(I+\delta)^a) &&
	\hat{\dot{V}}_{loss}:= \dot{V}_{loss}(t_{min})=
	\begin{cases}
		\frac{m_1}{\sqrt{t_{min}}}, &for~1\alpha\\
		\frac{r'd'}{r't_{min}+d'},&for~1\beta
		\end{cases}
\end{align}
The Lipschitz constant is then defined follows: 
\begin{align}
    K_{Lipschitz} := \max\{1,K_1,\dots,K_5\}
\end{align}
Evidently, the $\hat{\dot{V}}_{loss}$ in 1$\alpha$ is not defined at $t_{min}=0$. This is not ideal since it ties into the fact that when using 1$\alpha$ our system is not locally Lipschitz at $t=0$. This however is not a problem with the formula in 1$\beta$. Thus when using 1$\beta$ $X$ is already continuously dependent on $(t_0,X_0,f)$ and therefore quite stable under small errors.

\section{Derivation of surface area}
\label{sec:LSA_deriv}
In order to describe the crystal surface area $L_{SA}$ as a function of time, we aim to relate it to the crystal width. We assume that the crystal may be approximated by a prism with regular n-sided base.
We set \(r_i\) to be the height of the \(n\) triangles that form the base of the structure, such that \(2r_i = C_{width}(t)\). This holds true because $r_i$ is also the radius of the inner circle of each of these base triangles. Additionally, let \(h\) be the height of the prism and \(a\) the length of one side of the \(n\)-sided base. The surface area \(L_{SA}\) is given by:

\begin{align}
    L_{SA} = 2 \cdot B_A + a \cdot h \cdot n = n \cdot a \cdot r_i + a \cdot h \cdot n = n \cdot a \cdot \frac{C_{width}(t)}{2} + a \cdot h \cdot n.
\end{align}

Substituting \(a = C_{width}(t) \tan \left(\frac{\pi}{n}\right)\), we get:

\begin{align}
    L_{SA} = \frac{1}{2} \cdot n \cdot C_{width}(t)^2 \cdot \tan \left(\frac{\pi}{n}\right) + C_{width}(t) \cdot \tan \left(\frac{\pi}{n}\right) \cdot h \cdot n.
\end{align}

We define \(k_9 = \frac{1}{2} \cdot n \cdot \tan\left(\frac{\pi}{n}\right)\) and \(k_{10} = \tan \left(\frac{\pi}{n}\right) \cdot h \cdot n\), and therefore obtain:

\begin{align}
    L_{SA} = C_{width}(t)^2 \cdot k_9 + C_{width}(t) \cdot k_{10}.
\end{align}

Because we assume that the crystal is only growing in width and that we may neglect the growth in length, we remove the contribution of the base faces to the active surface area and obtain the following formula:

\begin{align}
    L_{SA} = k_{10} \cdot C_{width}(t).
\end{align}

\section{Material dependent fitting of lattice spacing}
\label{sec:L_fitted}

\begin{figure}[H]
    \centering
    \begin{subfigure}{0.49\textwidth}
    \centering
        \includegraphics[width = \textwidth]{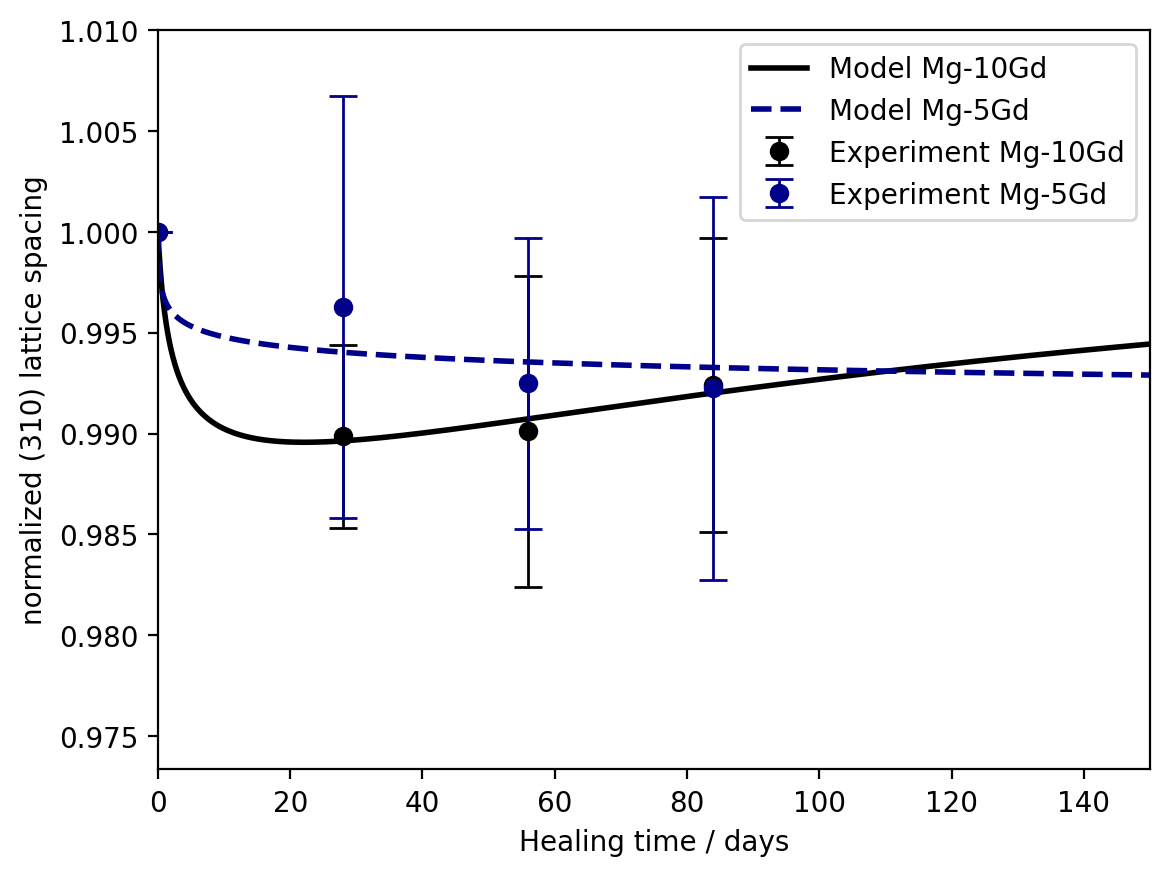}
        \caption{(310) lattice spacing $L_{310}$}
    \end{subfigure}
    \begin{subfigure}{0.49\textwidth}
    \centering
        \includegraphics[width = \textwidth]{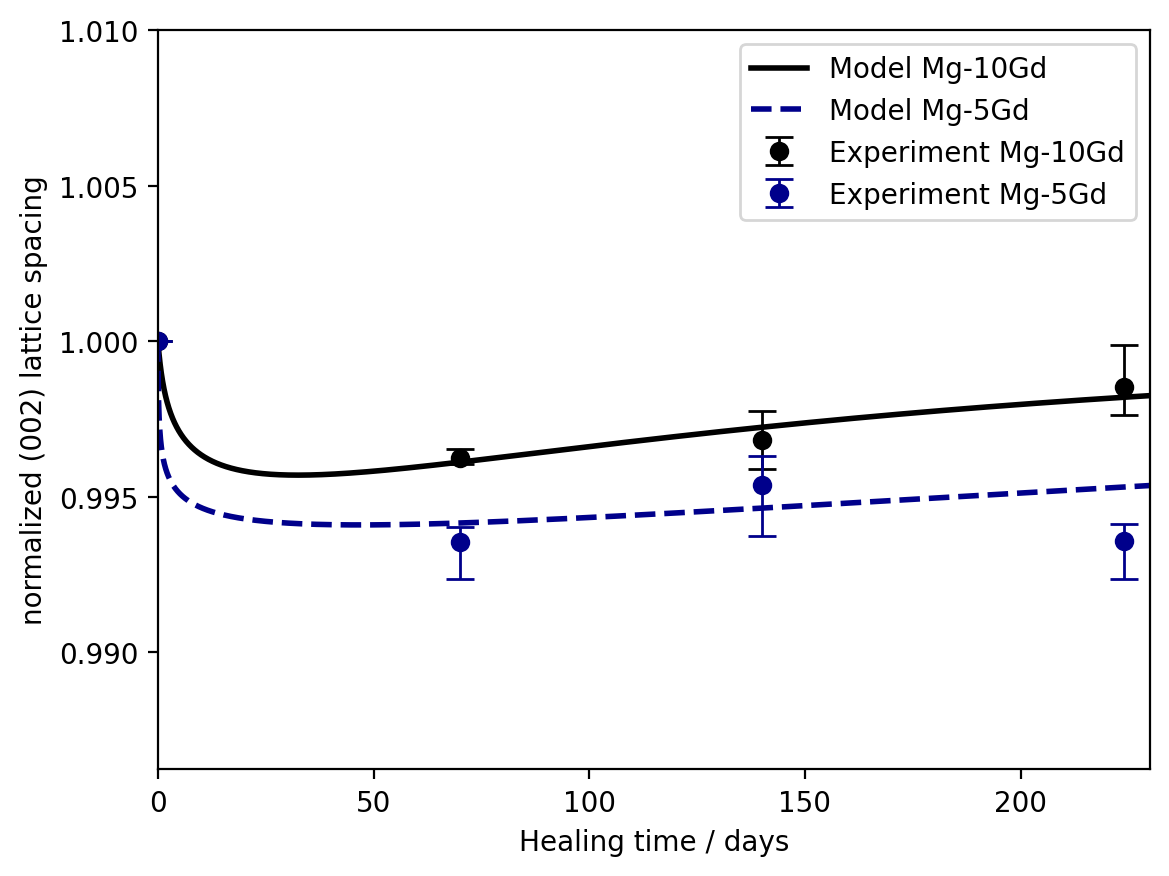}
        \caption{(002) lattice spacing $L_{002}$}
    \end{subfigure}
    \caption{Model output for (a) the normalized (310) and (b) the normalized (002) lattice spacing $L$ according to equation 1g for Mg-10Gd (black, solid line), and Mg-5Gd (blue, dashed line) implants. The experimental data in (a) is taken from \cite{ZELLERPLUMHOFF2020}, where it was given in mean $\pm$ standard deviation, while that in (b) is taken from \cite{IskhakovaCwieka2024}, which was given as median $\pm$ the 25/75th-percentile.}
    \label{fig:HAP_calboth}
\end{figure}

Figure~\ref{fig:HAP_calboth} displays the fitted lattice spacings and experimental data if the parameters in equation 1g are calibrated individually for both Mg-5Gd and Mg-10Gd. The NRMSE for the Mg-5Gd prediction is lowered by one order of magnitude (0.39 vs. 1.04 for (310) and 0.63 vs. 1.90 for (002)) when an individual fit is found and the MAE improved as well. However, in order to enable the kinetics shown by the data, $k_8$ decreases by one order of magnitude for the (002) fit and further to near zero for (310).



\end{document}